\documentclass[superscriptaddress,prd,aps,showpacs,nofootinbib,showkeywords,eqsecnum]{revtex4-2}
\usepackage{graphicx,color,amsmath,amsxtra}

\usepackage[final]{hyperref} 
\usepackage{epsf}
\usepackage{amssymb}
\usepackage{enumerate}
\usepackage{hhline}
\usepackage{array}
\usepackage{tabularx}
\usepackage{graphicx}                
\usepackage{epstopdf}
\usepackage{caption}
\usepackage{subcaption}
 \usepackage{multirow}
 \usepackage{booktabs}
 \usepackage{rotating} 
 \usepackage{cancel}

\begin{document}
\pagestyle{myheadings}

\title{A gap between two approaches of dimensional reduction for a six-dimensional Kaluza-Klein theory}
\author{Tuan Q. Do }
\email{tuan.doquoc@phenikaa-uni.edu.vn}
\affiliation{Phenikaa Institute for Advanced Study, Phenikaa University, Hanoi 12116, Vietnam}
\affiliation{Faculty of Basic Sciences, Phenikaa University, Hanoi 12116, Vietnam}
\author{W. F. Kao}
\email{homegore09@nycu.edu.tw}
\affiliation{
Institute of Physics, National Yang Ming Chiao Tung University, Hsin Chu 30010, Taiwan
}
\date{\today} 
\begin{abstract}
Inspired by the five-dimensional Kaluza-Klein theory, we would like to study the dimensional reduction issue of six-dimensional Kaluza-Klein extension in this paper. In particular, we will examine two possible approaches of dimensional reduction from six-dimensional spacetimes to four-dimensional ones. The first one is a direct dimensional reduction, i.e., from six-dimensional spacetimes directly to four-dimensional ones, via a $T^2\equiv S^1 \times S^1$ compactification, while the second one is an indirect dimensional reduction, i.e., from six-dimensional spacetimes to five-dimensional ones then four-dimensional ones, via two separated $S^1$ compactifications. Interestingly, we show that these two approaches lead to different four-dimensional effective actions although using the same six-dimensional metric. It could therefore address an important question of which approach is more reliable than the other. 
\end{abstract}
\maketitle
\section{Introduction} \label{intro}
Models of multiple fields, e.g., multiple scalar fields or multiple vector fields, have been investigated extensively in literature. Two main motivations for the introduction of multi-field cosmology are the cosmic inflation \cite{Wands:2007bd} and dark energy \cite{Copeland:2006wr,Eskilt:2022zky} issues, which have been among the hottest subjects of modern cosmology for recent decades. Understanding the origin of four-dimensional ($4D$) models of multiple fields is therefore important. For example, ones could think of scenarios that string theories and other higher-dimensional theories induce $4D$ effective actions with many moduli fields describing the higher dimensional degrees of freedom \cite{Wands:2007bd}.

It turns out that inducing higher dimensional theories/models to $4D$ ones requires reasonable and compatible mechanisms \cite{Quevedo:2010ui}. One of such mechanisms comes from the Kaluza-Klein (KK) theory \cite{Kaluza,Klein:1926tv}. The KK theory was first proposed by Kaluza \cite{Kaluza} with an effort to unify gravity to electromagnetism through the introduction of an extra dimension (the fifth dimension). Soon after the publication of Kaluza, Klein provided a reasonable mechanism based on a $S^1$ compactification condition, which is lately called the Klein's compactification mechanism, for the extra dimension \cite{Klein:1926tv}. It has had a long rich history and played a very important role not only in the high energy physics but also in the modern cosmology, e.g., see Refs. \cite{Witten:1981me,Salam:1981xd,Freund:1982pg,Chodos:1979vk,Cho:1986kd,Cvetic:1995sz,Kao:2000yw,Wesson,Cremmer:1978km,Freund:1980xh,Nastase:1999cb,Nastase:1999kf,Bailin:1987jd,Overduin:1997sri,Duff:1994tn,pope} for an incomplete list of literature. One remarkable example can be shown among important implications mentioned above is that the KK dimensional reduction has played an important role in the eleven-dimensional ($11D$) supergravity \cite{Cremmer:1978km,Freund:1980xh}. Recently, it has been shown that the KK dimensional reduction can helps to successfully construct $4D$ effective Lagrangians having potentials from eleven-dimensional supergravity ones \cite{Nastase:1999cb,Nastase:1999kf}.

Historically, Pauli is perhaps the first person developing the first consistent generalization of the five-dimensional ($5D$) KK theory to a higher dimensional internal space (in this case it is a two-dimensional internal space), according to Ref. \cite{Straumann:2000zc}. In particular, Pauli considered in his unpublished paper a six-dimensional ($6D$) KK theory and arrived through a dimensional reduction at the essentials of an $SU(2)$ gauge theory as stated in Ref. \cite{Straumann:2000zc}. Other interesting proposal on $6D$ extensions of the KK theory with a spontaneous compactification $S^2$ can be found in Refs. \cite{Randjbar-Daemi:1982opc}. See also Refs. \cite{Denef:2007pq,Seo:2023fuj} for recent related discussions. Motived by these works, we would also like to study in this paper a $6D$ extension of the KK theory along with two possible approaches of dimensional reduction. Our analysis is based on the interesting works on the $5D$ KK theory done in Refs. \cite{Thirring,Ichinose:2002kg}. In harmony with the prediction of multi scalar and gauge fields due to a dimensional reduction done in the seminal paper in Ref. \cite{Freund:1980xh},  these approaches all lead to $4D$ effective models of multi scalar and vector fields, whose rich cosmological implications have been explored extensively, e.g., see Refs. \cite{Chimento:2008ws,vandeBruck:2009gp,Saridakis:2009jq,Do:2017rva,Do:2023mqe} for an incomplete list of literature.
 
 It is worth noting that choosing a suitable dimensional reduction for higher-than-five-dimensional KK theories is an important issue, which has been discussed extensively by many people, e.g., see Ref. \cite{Cvetic:2000dm} for one of the most relevant literatures. In Ref. \cite{Cvetic:2000dm}, the authors have considered several dimensional reduction approaches based on $S^n$ (other than $T^n$) compactification for the $11D$ supergravity. In our current paper, the first approach of dimensional reduction is based on a 2-torus, i.e., $T^2 \equiv S^1 \times S^1$, compactification, by which we can directly reduce a pure $6D$ Ricci scalar to a $4D$ Ricci scalar plus pure kinetic terms of two scalar and two vector fields, in addition to a mixed kinetic term of two scalar fields.  We call this approach a direct $6D \to 4D$ reduction for convenience. See Refs. \cite{Duff:1994tn,pope} for extensive discussions on $n$-torus compactifications for $4+n$ dimensional spacetimes.  The second approach of dimensional reduction is based on the $S^1$ compactification, which was firstly introduced by Klein for the $5D$ Kaluza theory \cite{Klein:1926tv}. Since we will perform two $S^1$ compactifications separately, we will therefore call this approach an indirect $6D \to 5D \to 4D$ reduction.  The first $S^1$ compactification will help us to reduce a $6D$ Ricci scalar to the corresponding $5D$ Ricci scalar plus kinetic terms of one scalar and one vector fields. Then, the next  $S^1$ compactification  will lead this $5D$ Ricci scalar to the corresponding $4D$ Ricci scalar plus extra kinetic terms of another scalar and another vector fields. However, the resulted $4D$ effective action is different from that derived via the first approach of dimensional reduction. This means that there exists a significant gap between these two approaches.  In addition, it will be shown that the latter $S^1$ compactification cannot be possible if at least one field coming from the first $S^1$ compactification depends on the fifth coordinate. One could therefore ask an important question that which approach is more reliable than the other.

In summary, this paper will be organized as follows: (i) A brief introduction has been presented in Sec. \ref{intro}. (ii) A direct  $6D \to 4D$ reduction will be done in Sec. \ref{sec2}. (iii) An indirect $6D \to 5D \to  4D$ reduction will be performed in  Sec. \ref{sec3}. (iv) Concluding remarks will be mentioned in Sec. \ref{final}. (v) Additionally, related calculations for the results shown in Sec. \ref{sec2} will be listed in the Appendix.

{\bf Notations}: Before going to the next sections, it would be useful if we are clear about the notations used in these sections. This would avoid any misunderstandings so far. Objects having ``hat'' would be associated with $6D$ spacetimes;  Objects having ``tilde'' would be associated with $5D$ spacetimes; and objects having no ``hat'' and ``tilde'' would be associated with $4D$ spacetimes.  Roman (letters) indices  will be associated with general coordinates; while Greek (letters) indices will be associated with local Lorentz (Minkowski) coordinates. 
\section{Direct $6D \to 4D$ reduction}  \label{sec2}
Following Refs. \cite{Thirring,Ichinose:2002kg}, we would like to propose a $6D$ extension of the Kaluza-Klein theory \cite{Kaluza,Klein:1926tv} by introducing the corresponding metric,
\begin{align} \label{metric-1}
ds^2 = \hat g_{mn} dX^m dX^n =g_{ab} (x) dx^a dx^b +e^{2\sigma_1 (x)} \left[dy^1 -f_1 A_a^1 (x)dx^a \right]^2+e^{2\sigma_2 (x)} \left[dy^2 -f_2 A_a^2 (x)dx^a \right]^2,
\end{align}
where $y^1$ and $y^2$ stand for the fifth and sixth dimensions, respectively. Additionally, $g_{ab}$ is the $4D$ metric with $a, ~b=0,~1,~2,$ and $3$; while $X^m = (x^a,y^1,y^2)$ with $m=0,~1,~2,~3, ~5,$ and $6$. It is noted that $\sigma_1(x)$ and $\sigma_2(x)$ are the scalar (a.k.a. dilaton) fields, while $A_a^1(x)$ and $A_a^2(x)$ act as the $U(1)$ fields. All these fields depend only on $x^a$, not $y^1$ and $y^2$. Any objects carrying the ``hat''  should be understood as the $6D$ ones. Finally, $f_1$ and $f_2$ are just coupling constants. It is noted that various choices of higher dimensional metrics have also been proposed, e.g., see Ref.  \cite{Chodos:1979vk} for a choice ensuring a minimal coupling between scalar fields and the $4D$ Ricci scalar, or see Ref. \cite{Cho:1986kd} as well as Ref. \cite{Cvetic:1995sz} for other choices.

 It should be noted that the choice of above metric corresponds to the zeroth mode, i.e., $k_1=k_2=0$ and massless mode, of the Fourier expansions \cite{Klein:1926tv,Bailin:1987jd,Overduin:1997sri,Ichinose:2002kg},
\begin{align} 
 g_{ab}(X)=&\sum_{k_1,k_2 \in Z}  g^{(k_1,k_2)}_{ab}(x) e^{i k_1 \mu_1 y^1}e^{i k_2 \mu_2 y^2},\nonumber\\
A^1_a(X)=&\sum_{k_1,k_2 \in Z}  A^{1(k_1,k_2)}_{a}(x) e^{i k_1 \mu_1 y^1}e^{i k_2 \mu_2 y^2}, \quad A^2_a(X)=\sum_{k_1,k_2 \in Z}  A^{2(k_1,k_2)}_{a}(x) e^{i k_1 \mu_1 y^1}e^{i k_2 \mu_2 y^2},\nonumber\\
\sigma_1(X)=&\sum_{k_1,k_2 \in Z}  \sigma_1^{(k_1,k_2)}(x) e^{i k_1 \mu_1 y^1}e^{i k_2 \mu_2 y^2}, \quad \sigma_2(X)=\sum_{k_1,k_2 \in Z}  \sigma_2^{(k_1,k_2)}(x) e^{i k_1 \mu_1 y^1}e^{i k_2 \mu_2 y^2},
\end{align}
which are due to the $T^2 \equiv S^1 \times S^1$ compactification condition (which is an extension of  the $S^1$ Klein's compactification mechanism for $5D$ spacetimes) for two extra dimensions such as
\begin{align} \label{T2-condition}
 g_{ab}(X)&=\hat g_{mn} \left(x,y^1+\frac{2\pi}{\mu_1},y^2+\frac{2\pi}{\mu_2} \right),\\
A^1_a(X) &=A^1_a\left(x,y^1+\frac{2\pi}{\mu_1},y^2+\frac{2\pi}{\mu_2} \right),\quad A^2_a(X)  =A^2_a\left(x,y^1+\frac{2\pi}{\mu_1},y^2+\frac{2\pi}{\mu_2} \right),\\
\sigma_1(X)&=\sigma_1\left(x,y^1+\frac{2\pi}{\mu_1},y^2+\frac{2\pi}{\mu_2} \right),\quad \sigma_2(X)=\sigma_2\left(x,y^1+\frac{2\pi}{\mu_1},y^2+\frac{2\pi}{\mu_2} \right).
\end{align}
Here $\mu_1^{-1}$ and $\mu_2^{-1}$ are the compactification radius of the fifth and sixth dimensions, respectively. It is noted that only the massless $k_1=k_2=0$ mode is observable as required in the Kaluza's theory \cite{Bailin:1987jd,Overduin:1997sri}.

Now, we would like to use the Cartan formalism, whose basic details can be found in Refs. \cite{Ichinose:2002kg,Nakahara:2003nw}, to compute the corresponding geometric quantities such as the Ricci scalar. A reason for this choice is that the Cartan  formalism seems to be more simpler than the usual one, i.e., direct calculations of the Ricci scalar via the Christoffel symbols \cite{Thirring}. It should be noted that we have partially followed the conventions used in Ref. \cite{Ichinose:2002kg} for straightforward comparisons. 

First, we introduce the basis $\{\hat\theta^\mu\}$ of the cotangent manifold $(T^\ast_p M)$ as follows
\begin{equation} \label{metric-2}
ds^2 =\hat\theta^\mu \hat\theta^\nu \hat\eta_{\mu\nu},
\end{equation}
where $\hat\eta_{\mu\nu} ={\rm diag}(-1,1,1,1,1,1)$ and $\mu, ~\nu=0,~1,~2,~3,~\bar 5$, and $\bar 6$ are the local Lorentz (Minkowski) indices, which should be raised and lowered by $\hat\eta_{\mu\nu}$.   
In this section, note again that the Roman letters $m$ and $n$ have been used for the general coordinates, and therefore run as $m,~n=0,~1,~2,~3,~5$, and $6$, and will be raised and lowered by $\hat g_{mn}$, in contrast to the Greek letters $\mu$ and $\nu$. 
By definition, $\{\hat\theta^\mu  \equiv \hat e^\mu{}_m dX^m \}$
 is nothing but the dual of the basis $\{\hat e^\mu \equiv \hat e^{\mu m} \partial/\partial X^m \}$ of the tangent manifold $(T_p M)$, i.e.,
 \begin{equation} \label{eq1}
\langle \hat e^\mu, \hat\theta_\nu \rangle = \delta^\mu_\nu.
\end{equation}
Here, $\hat e^\mu{}_m$ is called {\it funfbein} (sometimes it is called {\it vielbeins} for the higher-than-four-dimensional spacetimes, e.g., see Ref. \cite{Nakahara:2003nw}). 

As a result, Eqs. \eqref{metric-1} and \eqref{metric-2} imply that
\begin{align} \label{def-hat-theta}
\hat\theta^{\bar 5} = e^{\sigma_1} \left(dy^1 -f_1 A_a^1 dx^a \right), \quad \hat\theta^{\bar 6} = e^{\sigma_2} \left(dy^2 -f_2 A_a^2 dx^a \right).
\end{align}
On the other hand, it turns out from the definition, $\hat\theta^\mu  \equiv \hat e^\mu{}_m dX^m$, that
\begin{align} \label{hat-theta-1}
\hat\theta^{\alpha} &=  e^\alpha{}_a dX^a +  \hat e^\alpha{}_5dX^5 + \hat e^\alpha{}_6 dX^6,\\
\label{hat-theta-2}
\hat\theta^{\bar 5} &= \hat e^{\bar 5}{}_a dX^a + \hat e^{\bar 5}{}_5 dX^5  + \hat e^{\bar 5}{}_6 dX^6,\\
\label{hat-theta-3}
\hat\theta^{\bar 6} &= \hat e^{\bar 6}{}_a dX^a + \hat e^{\bar 6}{}_5 dX^5  + \hat e^{\bar 6}{}_6 dX^6,
\end{align}
where $\alpha =0,~1,~2,~3$, and $\hat e^\alpha{}_a \equiv e^\alpha{}_a$ as the $4D$ part.  It is noted that the first Greek letters as $\alpha$, $\beta$, $\gamma$, and $\delta$ should be understood as the $4D$ index and therefore run from $0$ to $3$. Comparing these results with Eqs. \eqref{metric-1}, \eqref{metric-2}, and \eqref{def-hat-theta}, we arrive at
\begin{align}
&\hat e^\alpha{}_5 =  \hat e^\alpha{}_6= 0,\\
&\hat e^{\bar 5}{}_a = -f_1 e^{\sigma_1} A_a^1, \quad \hat e^{\bar 5}{}_5=e^{\sigma_1},\quad \hat e^{\bar 5}{}_6=0,\\
&\hat e^{\bar 6}{}_a = -f_2 e^{\sigma_2} A_a^2, \quad \hat e^{\bar 6}{}_5=0,\quad \hat e^{\bar 6}{}_6=e^{\sigma_2}. \label{hat-e-3}
\end{align}
For convenience,  $\hat e^\mu{}_m$ is written as
\begin{align}
\hat e^\mu{}_m = \left( {\begin{array}{*{20}c}
   {e^\alpha{}_a} & {0} & {0 }  \\
   {-f_1 e^{\sigma_1} A_a^1 } & {e^{\sigma_1}} & { 0}  \\
     {-f_2 e^{\sigma_2} A_a^2 } & {0 } & {e^{\sigma_2} } \\
 \end{array} } \right).
\end{align}
Hence, its inverse, which must satisfy the following constraints, $\hat e^\mu{}_m \hat e_\mu{}^n=\delta _m^n$ and $\hat e^\mu{}_m \hat e_\nu{}^m=\delta^\mu_\nu$, can be defined to be
\begin{align}
\hat e_\mu{}^m = \left( {\begin{array}{*{20}c}
   {e_\alpha{}^a} & {0} & {0 }  \\
   {f_1 A_\alpha^1 } & {e^{-\sigma_1}} & { 0}  \\
     {f_2 A_\alpha^2 } & {0 } & {e^{-\sigma_2} } \\
 \end{array} } \right),
\end{align}
where $A_\alpha^1 = e_\alpha{}^a A_a^1$ and $A_\alpha^2 = e_\alpha{}^a A_a^2$.

Now, we will consider the corresponding Cartan's structure equations, which encode all properties of the models. The first one for the torsionless case, i.e., $\hat T^\mu =0$, is given by \cite{Ichinose:2002kg}
\begin{equation} \label{Cartan-equation-1}
d\hat\theta^\mu +\hat\omega^\mu{}_\nu \wedge \hat\theta^\nu = \hat T^\mu =0,
\end{equation}
where $\hat\omega^\mu{}_\nu $ is the connection 1-form, which satisfies the following anti-symmetric property, $\hat\omega_{\mu\nu} =-\hat\omega_{\nu\mu}$. In addition, $d\hat\theta^\mu$ should not be understood as an ordinary derivative , but an exterior derivative \cite{Deruelle:2018ltn}. One can therefore derive all components of $\hat\omega^\mu{}_\nu$ from this equation. As a result, we have the following equation for the $\hat\theta^{\bar 5}$ as
\begin{align}
d\hat\theta^{\bar 5} +\hat\omega^{\bar 5}{}_\nu \wedge \hat\theta^\nu &=0,
\end{align}
which can be expanded to be
\begin{align}
 -f_1 \partial_b \left(e^{\sigma_1}A^1_a \right) dX^b \wedge dX^a + \partial_b e^{\sigma_1} dX^b \wedge dX^5 + \hat\omega^{\bar 5}{}_\alpha \wedge \left( e^\alpha{}_a dX^a \right) +\hat\omega^{\bar 5}{}_{\bar 5} \wedge \left(-f_1 e^{\sigma_1} A^1_c dX^c + e^{\sigma_1} dX^5 \right)  =0,
\end{align}
with the help of Eqs. \eqref{hat-theta-1} and \eqref{hat-theta-2}.  It is noted that we have set $\hat\omega^{\bar 5}{}_{\bar 6} =0$ due to the fact that there are no other terms containing $\hat \theta^{\bar 6}$. It turns out, due to the anti-symmetric property of $\hat\omega_{\mu\nu}$, that 
\begin{equation}
\hat\omega^{\bar 5}{}_{\bar 5} = 0,
\end{equation}
then we have from the above equation that 
\begin{align}
\hat\omega^{\bar 5}{}_\alpha e^\alpha{}_a = \left( \partial_a\sigma_1 \right)e^{\sigma_1} \left(dX^5-f_1 A_b^1 dX^b \right) - \frac{1}{2}f_1 e^{\sigma_1} \left(\partial_a A_b^1 - \partial_b A_a^1 \right) dX^b.
\end{align}
Thanks to Eqs. \eqref{hat-theta-1} and \eqref{hat-theta-2}, we will simplify this equation  as
\begin{align}
\hat\omega^{\bar 5}{}_\alpha =  \left( \partial_a\sigma_1 \right) e_\alpha{}^a \hat\theta^{\bar 5} - \frac{1}{2} f_1 e^{\sigma_1} F^1_{\alpha\beta} \hat\theta^\beta
\end{align} 
for convenience. Here, $F_{\alpha\beta}^1= e_\alpha{}^a e_\beta{}^b F^1_{ab} $ and $F^1_{ab}=\partial_a A_b^1 -\partial_bA_a^1$. Similarly, we have for the $\hat\theta^{\bar 6}$ that
\begin{align}
\hat\omega^{\bar 6}{}_{\bar 6} = 0,\quad \hat\omega^{\bar 6}{}_\alpha =  \left( \partial_a\sigma_2 \right) e_\alpha{}^a \hat\theta^{\bar 6} - \frac{1}{2} f_2 e^{\sigma_2} F^2_{\alpha\beta} \hat\theta^\beta,
\end{align} 
where  $F_{\alpha\beta}^2= e_\alpha{}^a e_\beta{}^b F^2_{ab} $ and $F^2_{ab}=\partial_a A_b^2 -\partial_bA_a^2$. It should be noted that $\hat\omega^{\bar 6}{}_{\bar 5} =0$, similar to $\hat\omega^{\bar 5}{}_{\bar 6}$. 

So far, $\hat\omega^{\bar 5}{}_{\alpha}$, $\hat\omega^{\bar 5}{}_{\bar 5}$, $\hat\omega^{\bar 6}{}_{\alpha}$, and $\hat\omega^{\bar 6}{}_{\bar 6}$ have been worked out. The next geometrical object needs to be revealed is the remaining $\hat\omega^{\alpha}{}_{\beta}$ one.  As a result, this can be done by considering the following equation of $\hat\theta^{\alpha}$ given by
\begin{align} \label{first-cartan-equation-3}
d\hat\theta^\alpha +\hat\omega^\alpha{}_\nu \wedge \hat\theta^\nu &=0
\end{align}
or equivalently,
\begin{align}
 \partial_b \left(e^\alpha{}_a\right) dX^b \wedge dX^a + \hat \omega^\alpha{}_\beta \wedge \hat\theta^\beta + \hat \omega^\alpha{}_{\bar 5} \wedge \hat\theta^{\bar 5} +\hat \omega^\alpha{}_{\bar 6} \wedge \hat\theta^{\bar 6}&=0.
\end{align}
For convenience, we decompose an effective $4D$ connection such as
\begin{equation} \label{decompose}
\hat\omega^\alpha{}_\beta =\omega^\alpha{}_\beta + \bar\omega^\alpha{}_\beta,
\end{equation}
 and we therefore rewrite Eq. \eqref{first-cartan-equation-3} as follows
\begin{align} \label{first-cartan-equation-4}
&\partial_b \left(e^\alpha{}_a\right) dX^b \wedge dX^a + \omega^\alpha{}_\beta \wedge \hat\theta^\beta +\bar\omega^\alpha{}_\beta \wedge \hat\theta^\beta+ \hat \omega^\alpha{}_{\bar 5} \wedge \hat\theta^{\bar 5} +\hat \omega^\alpha{}_{\bar 6} \wedge \hat\theta^{\bar 6}=0.
\end{align}
It turns out that the first two terms in this equation form the following $4D$ Cartan's structure equation, in which $\omega^\alpha{}_\beta$ acts as the pure $4D$ connection,
\begin{equation}
d\theta^\alpha +\omega^\alpha{}_\beta \wedge \theta^\beta=0,
\end{equation}
where $\hat\theta^\alpha \equiv \theta^\alpha$ as an $4D$ basis.
Hence, Eq. \eqref{first-cartan-equation-4} simply reduces to
\begin{align} \label{first-cartan-equation-5}
\bar\omega^\alpha{}_\beta \wedge \hat\theta^\beta+ \hat \omega^\alpha{}_{\bar 5} \wedge \hat\theta^{\bar 5} +\hat \omega^\alpha{}_{\bar 6} \wedge \hat\theta^{\bar 6}=0.
\end{align}
Given $\hat\omega^{\bar 5}{}_{\alpha} $ and $\hat\omega^{\bar 6}{}_{\alpha} $ defined above, we able to obtain that 
\begin{align}
\hat\omega^\alpha{}_{\bar 5} = \hat\eta^{\alpha \delta} \hat\eta_{\bar 5\bar 5} \hat\omega_\delta{}^{\bar 5},\quad \hat\omega^\alpha{}_{\bar 6}  = \hat\eta^{\alpha \delta} \hat\eta_{\bar 6 \bar 6}\hat\omega_\delta{}^{\bar 6},
\end{align}
where $\hat\omega_\delta{}^{\bar 5} = -\hat\omega^{\bar 5}{}_\delta$ and $\hat\omega_\delta{}^{\bar 6} = -\hat\omega^{\bar 6}{}_\delta$. Now, it is straightforward to obtain 
\begin{equation}
\bar\omega^\alpha{}_\beta = \frac{1}{2}f_1 e^{\sigma_1} F^{1\alpha}{}_\beta \hat\theta^{\bar 5}  +\frac{1}{2}f_2 e^{\sigma_2} F^{2\alpha}{}_\beta \hat\theta^{\bar 6},
\end{equation}
provided that $\hat\theta^{\bar 5} \wedge \hat\theta^{\bar 5} = \hat\theta^{\bar 6} \wedge \hat\theta^{\bar 6}  =0$. Finally, we are able to write, according to Eq. \eqref{decompose}, the explicit definition of the effective $4D$ connection,
\begin{equation}
\hat\omega^\alpha{}_\beta = \omega^\alpha{}_\beta + \frac{1}{2}f_1 e^{\sigma_1} F^{1\alpha}{}_\beta \hat\theta^{\bar 5}  +\frac{1}{2}f_2 e^{\sigma_2} F^{2\alpha}{}_\beta \hat\theta^{\bar 6}.
\end{equation}

Now, we consider the second Cartan's structure equation, which is given by \cite{Ichinose:2002kg}
\begin{equation} \label{second}
d\hat\omega_{\mu\nu} +\hat\omega_{\mu \sigma} \wedge \hat\omega^\sigma{}_\nu =\frac{1}{2} \hat R_{\mu\nu\sigma \tau} \hat\theta^\sigma \wedge \hat\theta^\tau.
\end{equation}
Thanks to the definitions of the connection 1-form $\hat\omega^\mu{}_\nu$ defined above, we are able to derive all the corresponding non-vanishing components of $6D$ Riemann tensor $\hat R_{\mu\nu\sigma \tau}$ (see the Appendix \ref{app1} for detailed calculations).  Consequently, we are able to determine the following $6D$ Ricci scalar as (see the Appendix \ref{app2} for a complete list of non-vanishing relevant components of $6D$ Ricci tensor $\hat R_{\mu\nu}$),
\begin{align}
\hat R =&~ \hat \eta^{\alpha\beta}  \hat R_{\alpha\beta} +\hat \eta^{\bar 5 \bar 5}  \hat R_{\bar 5 \bar 5}+\hat \eta^{\bar 6 \bar 6}  \hat R_{\bar 6\bar 6} \nonumber\\
=&~ R - \frac{1}{4} {f_1^2} e^{2\sigma_1} F^{1\alpha \beta} F^1_{\alpha \beta}  - \frac{1}{4} {f_2^2} e^{2{\sigma_2}} F^{2\alpha \beta} F^2_{\alpha \beta} \nonumber\\
&  - 2 \left(\partial_a \sigma_1 \right) \left(\partial^a \sigma_1 \right) -  2 {D^2 \sigma_1} - 2  \left(\partial_a \sigma_2 \right) \left(\partial^a \sigma_2 \right) -2 {D^2 \sigma_2} -2 \left(\partial_a \sigma_1 \right) \left(\partial^a\sigma_2 \right),
\end{align}
where $D_\mu$ is the covariant derivative. This result, especially the coefficients of kinetic terms and the existence of mixed kinetic term, can be easily verified in the flat limit of $g_{ab}$, i.e., $g_{ab} = \eta_{ab}$.

It should be noted that $\sqrt{-\hat g}= e^{\sigma_1 +\sigma_2} \sqrt{-g} $. Therefore, we arrive at the following result,
\begin{align} \label{4D-effective-action}
S = \frac{1}{16\pi \hat G}\int d^4 x dy^1 dy^2 \sqrt{- \hat g} \hat R =&~\frac{1}{16\pi \hat G} \int d^4 x dy^1 dy^2 \sqrt{-g} e^{\sigma_1+\sigma_2 } \nonumber\\
& \times \left[ R - \frac{1}{4} {f_1^2} e^{2\sigma_1} F^{1\alpha \beta} F^1_{\alpha \beta}  - \frac{1}{4} {f_2^2} e^{2{\sigma_2}} F^{2\alpha \beta} F^2_{\alpha \beta} \right. \nonumber\\
&~ \left.  - 2  \left(\partial_a \sigma_1 \right) \left(\partial^a \sigma_1 \right) -2 {D ^2 \sigma_1} -2   \left(\partial_a \sigma_2 \right) \left(\partial^a \sigma_2 \right) - 2 {D ^2 \sigma_2} -2   \left(\partial_a \sigma_1 \right) \left(\partial^a\sigma_2 \right) \right],
\end{align}
where $\hat G$ is a $6D$ gravitational constant, similar to a $5D$ gravitational constant of the Kaluza-Klein theory \cite{Bailin:1987jd,Overduin:1997sri}.
Furthermore, using the $T^2$ compactification mechanism, we are able to define an effective $4D$ gravitational constant $G$ as follows
\begin{equation}
G= \frac{\hat G}{\int dy^1 dy^2} = \frac{\hat G \mu_1 \mu_2}{4\pi^2}.
\end{equation} 
Hence, we can rewrite the above action as
\begin{align}
S=  \int d^4 x \sqrt{-g} e^{\sigma_1+\sigma_2 } &\left[ \frac{R}{2\kappa} - \frac{1}{4} {f_1^2} e^{2\sigma_1} F^{1\alpha \beta} F^1_{\alpha \beta}  - \frac{1}{4} {f_2^2} e^{2{\sigma_2}} F^{2\alpha \beta} F^2_{\alpha \beta} - \frac{1}{\kappa}  \left(\partial_a \sigma_1 \right) \left(\partial^a \sigma_1 \right)  \right. \nonumber\\
&~ \left.  - \frac{1}{\kappa} {D ^2 \sigma_1} - \frac{1}{\kappa}\left(\partial_a \sigma_2 \right) \left(\partial^a \sigma_2 \right) - \frac{1}{\kappa} {D ^2 \sigma_2} - \frac{1}{\kappa}  \left(\partial_a \sigma_1 \right) \left(\partial^a\sigma_2 \right) \right],
\end{align}
with $\kappa \equiv 8\pi G$ as usual and the re-scaling such as $ A^1_a/\sqrt{2\kappa} \to A^1_a$ and $ A^2_a/\sqrt{2\kappa} \to A^2_a$ has been used.

Furthermore, we can reduce this action to the following one via the well-known   integration by parts,
\begin{align} \label{Jordan}
S= \int d^4 x \sqrt{-g} e^{\sigma_1+\sigma_2 }&  \left[ \frac{R}{2\kappa} - \frac{1}{4} {f_1^2} e^{2\sigma_1} F^{1\alpha \beta} F^1_{\alpha \beta}  - \frac{1}{4} {f_2^2} e^{2{\sigma_2}} F^{2\alpha \beta} F^2_{\alpha \beta}  +\frac{1}{\kappa} \left(\partial_a \sigma_1 \right) \left(\partial^a\sigma_2 \right) \right].
\end{align}

It appears that this action is written in the Jordan frame, in which the  scalar (dilaton) fields, $\sigma_1$ and $\sigma_2$, are non-minimally coupled to the Ricci scalar. It is well known that this action can be reduced to one written in the Einstein frame, in which $\sigma_1$ and $\sigma_2$ are minimally coupled to the Ricci scalar,  by using the standard technique, i.e., the conformal transformation, e.g., see Refs. \cite{Overduin:1997sri,pope,Dabrowski:2008kx} for example. More interestingly, this conformal transformation could be useful to fix the wrong sign of the kinetic energy of scalar fields. As a result, we would like to take the corresponding conformal transformation 
\begin{equation}
\bar g_{ab} = \Omega^2(x) g_{ab},
\end{equation}
with 
\begin{equation}
\Omega^2(x) =e^{\sigma_1(x)+\sigma_2(x)}.
\end{equation}
Consequently, the following quantities will transform accordingly as follows \cite{Dabrowski:2008kx}
\begin{align}
\sqrt{-g} &= e^{-2\sigma_1-2\sigma_2} \sqrt{-\bar g},\\
R &= e^{\sigma_1+\sigma_2} \left[ \bar R -\frac{3}{2} \bar\partial_a \left(\sigma_1 +\sigma_2\right) \bar\partial^a \left(\sigma_1 +\sigma_2 \right) \right],\\
\sqrt{-g} F^2 &= \sqrt{-\bar g} {\bar F}^2,\\
\partial_a \sigma_1 \partial^a \sigma_1 &= e^{\sigma_1 + \sigma_2} \bar \partial_a \sigma_1 \bar\partial^a \sigma_1,\\
\partial_a \sigma_2 \partial^a \sigma_2 &= e^{\sigma_1 + \sigma_2} \bar \partial_a \sigma_2 \bar\partial^a \sigma_2,\\
\partial_a \sigma_1 \partial^a \sigma_2 &= e^{\sigma_1 + \sigma_2} \bar \partial_a \sigma_1 \bar\partial^a \sigma_2.
\end{align}
Plugging these results into Eq. \eqref{Jordan}, we have the corresponding action in the Einstein frame,
\begin{align} \label{Einstein}
S=  \int d^4 x \sqrt{-\bar g} & \left[ \frac{\bar R}{2\kappa} - \frac{1}{4} {f_1^2} e^{3\sigma_1 +\sigma_2} {\bar F}^{1\alpha \beta} {\bar F} ^1_{\alpha \beta}  - \frac{1}{4} {f_2^2} e^{\sigma_1+ 3{\sigma_2}} {\bar F}^{2\alpha \beta} {\bar F}^2_{\alpha \beta} \right. \nonumber\\
& \left.  - \frac{3}{4\kappa} \left(\bar \partial_a \sigma_1 \right) \left(\bar \partial^a \sigma_1 \right)  - \frac{3}{4\kappa}  \left(\bar \partial_a \sigma_2 \right) \left(\bar \partial^a \sigma_2 \right) -\frac{1}{2\kappa} \left(\bar \partial_a \sigma_1 \right) \left(\bar \partial^a \sigma_2 \right) \right].
\end{align}
Now, we have an important result that all kinetic terms of dilaton fields have a correct sign in the Einstein frame. It turns out that the last term in this action is nothing but a mixed kinetic term of two scalar fields $\sigma_1$ and $\sigma_2$. It is worth noting that there have existed a number of papers in literature investigating a multi-field scenario with mixed kinetic terms of scalar fields, e.g., see Refs. \cite{Chimento:2008ws,vandeBruck:2009gp,Saridakis:2009jq,Do:2017rva}. Therefore, the present paper could provide a natural mechanism revealing the origin of model of multi-field scenario with mixed kinetic terms. Additionally, the present paper could also provide an additional natural mechanism for models of multi vector fields non-minimally coupled to multi scalar ones, e.g., see Ref. \cite{Do:2023mqe}.
\section{Indirect $6D \to 5D \to  4D$ reduction} \label{sec3}
So far, a direct reduction from $6D$ spacetimes to $4D$ ones via the $T^2 \equiv S^1 \times S^1$ compactification has been demonstrated explicitly. In this section, we would like to consider another possible approach of dimensional reduction, where  a dimensional gap is one, i.e., from $6D$ spacetimes to $5D$ ones, then from $5D$  spacetimes to $4D$ ones, via two separated $S^1$ compactifications. Therefore, the $T^2$ compactification conditions shown in Eq. \eqref{T2-condition} will no longer be valid in this section.  For convenience, we will call this approach an direct $6D \to 5D \to  4D$ reduction.  In this approach, the $6D$ metric \eqref{metric-1} should be replaced by another $6D$ one given by
\begin{align} \label{metric-3}
ds^2 &= \hat g_{mn} dX^m dX^n =\tilde g_{\tilde a\tilde b} (x,y^1) dX^{\tilde a} dX^{\tilde b} +e^{2\tilde  \sigma_2 (x,y^1)} \left[dy^2 -f_2 \tilde A_{\tilde a}^2 (x,y^1)dx^{\tilde a} \right]^2,
\end{align}
where $\tilde g_{\tilde a \tilde b}$ is the $5D$ metric with $\tilde a, ~\tilde b=0,~1,~2,~3,$ and $5$. It is noted that $\tilde \sigma_2$ and $\tilde A_{\bar m}^2$ are now assumed to be functions of not only $x$ but also the fifth dimension $y^1$ for a general consideration. Additionally, the remaining fields $\sigma_1$ and $A_1$ have been absorbed into the 5D metric $\tilde g_{\tilde a\tilde b}$ for convenience. As a result, the equation $ds^2 =\hat\theta^\mu \hat\theta^\nu \hat\eta_{\mu\nu}$ implies that 
\begin{equation} \label{def-hat-theta-new}
 \hat\theta^{\bar 6} = e^{\tilde \sigma_2} \left(dy^2 - f_2 \tilde A_{\tilde a}^2 dx^{\tilde a} \right).
\end{equation}
On the other hand, the definition $\hat\theta^\mu  \equiv \hat e^\mu{}_m dX^m$ implies that
\begin{align} \label{hat-theta-1-new}
\hat\theta^{\tilde \alpha} &=  \hat e^{\tilde\alpha}{}_{\tilde a} dX^{\tilde a}  + \hat e^{\tilde\alpha}{}_6 dX^6,\\
\label{hat-theta-3-new}
\hat\theta^{\bar 6} &= \hat e^{\bar 6}{}_{\tilde a} dX^{\tilde a}  + \hat e^{\bar 6}{}_6 dX^6,
\end{align}
where $\tilde \alpha =0,~1,~2,~3,$ and $\bar 5$.  It is noted again that the first Greek letters as $\alpha$, $\beta$, $\gamma$, and $\delta$ with a ``tilde'' should be understood as the $5D$ indices in this section and therefore run from $0$ to $\bar 5$. Comparing these results with Eqs. \eqref{metric-3}, \eqref{metric-2}, and \eqref{def-hat-theta-new}, we arrive at
\begin{align}
 \hat e^{\tilde \alpha}{}_6= 0,\quad \hat e^{\bar 6}{}_{\tilde a} = -f_2 e^{\tilde\sigma_2} \tilde A_{\tilde a}^2,\quad \hat e^{\bar 6}{}_6=e^{\tilde\sigma_2}. 
\end{align}
For convenience,  $\hat e^\mu{}_m$ is written as
\begin{align}
\hat e^\mu{}_m = \left( {\begin{array}{*{20}c}
   {\hat e^{\tilde\alpha}{}_{\tilde a}} & {0}   \\
     {-f_2 e^{\tilde\sigma_2} \tilde A_{\tilde a}^2 } &  {e^{\tilde\sigma_2} } \\
 \end{array} } \right).
\end{align}
Hence, its inverse can be defined to be
\begin{align}
\hat e_\mu{}^m = \left( {\begin{array}{*{20}c}
   {\hat e_{\tilde\alpha}{}^{\tilde a}} & {0}   \\
     {f_2 \tilde A_{\tilde\alpha}^2 } &  {e^{-\tilde\sigma_2} } \\
 \end{array} } \right),
\end{align}
where $\tilde A_{\tilde \alpha}^2 = \hat e_{\tilde\alpha}{}^{\tilde a} A_{\tilde a}^2$. Now, we consider the first Cartan’s structure equations,
\begin{equation} \label{Cartan-equation-1-new}
d\hat\theta^\mu +\hat\omega^\mu{}_\nu \wedge \hat\theta^\nu =0.
\end{equation}
For this case, the specific equation for the $\hat \theta^{\bar 6}$ turns out to be
\begin{equation}
d\hat\theta^{\bar 6} +\hat\omega^{\bar 6}{}_\nu \wedge \hat\theta^\nu =0,
\end{equation}
which can be expanded as follows
\begin{align} \label{theta-6-new}
 -f_2 \partial_{\tilde b} \left( e^{\tilde\sigma_2} \tilde A^2_{\tilde a} \right) dX^{\tilde b} \wedge dX^{\tilde a} + \partial_{\tilde b}  e^{\tilde\sigma_2} dX^{\tilde b}  \wedge dX^6 + \hat\omega^{\bar 6}{}_{\tilde\alpha} \wedge \left( \hat e^{\tilde\alpha}{}_{\tilde a} dX^{\tilde a} \right) +\hat\omega^{\bar 6}{}_{\bar 6} \wedge \left(-f_2  e^{\tilde\sigma_2} \tilde A^2_{\tilde a} dX^{\tilde a} +  e^{\tilde\sigma_2} dX^6 \right)  =0.
\end{align}
Following the previous section, we have 
\begin{equation}
\hat\omega^{\bar 6}{}_{\bar 6} =0,
\end{equation}
by which we will obtain from Eq. \eqref{theta-6-new} that
\begin{equation}
\hat\omega^{\bar 6}{}_{\tilde \alpha} \hat e^{\tilde\alpha}{}_{\tilde a}  = \left( \partial_{\tilde a } \tilde\sigma_2 \right) e^{\tilde\sigma_2} \left(dX^6 -f_2 \tilde A_{\tilde b}^2 dX^{\tilde b} \right)  - \frac{1}{2}f_2 e^{\tilde\sigma_2} \left(\partial_{\tilde a} \tilde A_{\tilde b}^2 - \partial_{\tilde b} \tilde A_{\tilde a}^2 \right) dX^{\tilde b}.
\end{equation}
Now, using Eqs. \eqref{hat-theta-1-new} and \eqref{hat-theta-3-new} we will rewrite the above expression as
\begin{equation}
\hat\omega^{\bar 6}{}_{\tilde\alpha} = \left( \partial_{\tilde a } \tilde\sigma_2 \right) \hat e_{\tilde\alpha}{}^{\tilde a} \hat \theta^{\bar 6} -\frac{1}{2}f_2 e^{\tilde\sigma_2} \tilde F^2_{\tilde\alpha\tilde \beta} \hat\theta^{\tilde\beta},
\end{equation}
where $\tilde F^2_{\tilde\alpha\tilde\beta} = \hat e_{\tilde\alpha}{}^{\tilde a} \hat e_{\tilde\beta}{}^{\tilde b} \tilde F^2_{\tilde a \tilde b}$ and $\tilde F^2_{\tilde a \tilde b} = \partial_{\tilde a} \tilde A_{\tilde b}^2 - \partial_{\tilde b} \tilde A_{\tilde a}^2 $.

So far, $\hat \omega^{\bar 6}{}_{\tilde\alpha}$ and  $\hat \omega^{\bar 6}{}_{\bar 6}$ have been determined. The remaining  $\hat \omega^{\tilde\alpha}{}_{\tilde\beta}$ will be figured out from  the following equation,
\begin{equation}
d\hat\theta^{\tilde\alpha} +\hat \omega^{\tilde\alpha}{}_\nu \wedge \hat\theta^\nu =0,
\end{equation}
which can be expanded explicitly as follows
\begin{align} \label{hat-theta-new} 
 \partial_{\tilde b} \left( \hat e^{\tilde\alpha}{}_{\tilde a}\right) dX^{\tilde b} \wedge dX^{\tilde a} + \hat \omega^{\tilde\alpha}{}_{\tilde\beta} \wedge \hat\theta^{\tilde\beta}  +\hat \omega^{\tilde\alpha}{}_{\bar 6} \wedge \hat\theta^{\bar 6}=0.
\end{align}
Using the useful technique mentioned in the previous section, we are going to decompose an effective $5D$ connection, $\hat\omega^\alpha{}_\beta$, into two parts such as
\begin{equation} \label{decompose-new}
\hat\omega^\alpha{}_\beta = \tilde \omega^\alpha{}_\beta + \bar{\bar\omega}^\alpha{}_\beta,
\end{equation}
where the first one, i.e., $ \tilde \omega^\alpha{}_\beta$ acting as the pure $5D$ connection in the $5D$ Cartan's structure  equation, 
\begin{equation} \label{5D-Cartan's-structure-equation}
d \hat \theta ^\alpha +  \tilde \omega^\alpha{}_\beta \wedge \hat\theta^\beta =0.
\end{equation}
As a result, Eq. \eqref{hat-theta-new} is rewritten as follows
\begin{equation}
 \partial_{\tilde b} \left(\hat e^{\tilde\alpha}{}_{\tilde a}\right) dX^{\tilde b} \wedge dX^{\tilde a} + \tilde \omega^{\tilde\alpha}{}_{\tilde\beta} \wedge \hat\theta^{\tilde\beta} +\bar{\bar \omega}^{\tilde\alpha}{}_{\tilde\beta} \wedge \hat\theta^{\tilde\beta}  +\hat \omega^{\tilde\alpha}{}_{\bar 6} \wedge \hat\theta^{\bar 6}=0,
 \end{equation}
 which is reduced to a simple one, thanks to Eq. \eqref{5D-Cartan's-structure-equation},
 \begin{equation}
 \bar{\bar \omega}^{\tilde\alpha}{}_{\tilde\beta} \wedge \hat\theta^{\tilde\beta}  +\hat \omega^{\tilde\alpha}{}_{\bar 6} \wedge \hat\theta^{\bar 6}=0.
 \end{equation}
This equation can be solved to give
\begin{equation}
\bar{\bar \omega}^{\tilde\alpha}{}_{\tilde\beta} =\frac{1}{2}f_2 e^{\tilde\sigma_2} \tilde F^{2\tilde\alpha}{}_{\tilde\beta} \hat\theta^{\bar 6}.
\end{equation}
Hence, it concludes that
\begin{equation}
\hat\omega^{\tilde\alpha}{}_{\tilde\beta} = \tilde \omega^{\tilde\alpha}{}_{\tilde\beta} +\frac{1}{2}f_2 e^{\tilde\sigma_2} \tilde F^{2\tilde\alpha}{}_{\tilde\beta} \hat\theta^{\bar 6}.
\end{equation}
So far, all components of $\hat \omega^\mu{}_\nu$ have been worked out. Given these results, we can move on to the second Cartan’s structure equation \eqref{second} to work out all non-vanishing components of the corresponding $6D$ Riemann tensor $\hat R_{\mu\nu\sigma\tau}$. Consequently,  the corresponding $6D$ Ricci scalar can be determined later.  After a bit lengthy calculations, we would arrive at the expected result,
\begin{equation} \label{6D-Ricci-new}
\hat R = \tilde R -\frac{1}{4}f_2^2 e^{2\tilde\sigma_2} \tilde F_{2\tilde\alpha\tilde\beta}\tilde F^2_{\tilde\alpha\tilde\beta}- 2  \left(\partial_{\tilde a} \tilde\sigma_2 \right) \left(\partial^{\tilde a} \tilde\sigma_2 \right) - 2{D^2 \tilde\sigma_2},
\end{equation}
here $\tilde R$ is nothing but a $5D$ pure Ricci scalar. Consequently, it appears that 
\begin{align} \label{5D-effective-action}
S = \frac{1}{16\pi \hat G}\int d^5 x  dy^2 \sqrt{- \hat g} \hat R=\frac{1}{16\pi \hat G} \int d^5 x dy^2 \sqrt{-\tilde g} e^{\tilde\sigma_2 } \left[ \tilde R   - \frac{1}{4} {f_2^2} e^{2{\tilde\sigma_2}} \tilde F^{2\tilde\alpha \tilde\beta} \tilde F^2_{\tilde\alpha \tilde\beta}  -2   \left(\partial_{\tilde a} \tilde\sigma_2 \right) \left(\partial^{\tilde a} \tilde\sigma_2 \right) -2 {D ^2 \tilde\sigma_2} \right],
\end{align}
which can be further $S^1$ compactified to be an $5D$ effective action,
\begin{align} \label{5D-effective-action-2}
S =\int d^5 x \sqrt{-\tilde g} e^{\tilde\sigma_2 } \left[ \frac{ \tilde R }{2 \tilde \kappa}  - \frac{1}{8 \tilde \kappa} {f_2^2} e^{2{\tilde\sigma_2}} \tilde F^{2 \tilde \alpha \tilde \beta} \tilde F^2_{\tilde\alpha \tilde \beta}  - \frac{1}{ \tilde \kappa}  \left(\partial_{\tilde a} \tilde\sigma_2 \right) \left(\partial^{\tilde a} \tilde\sigma_2 \right) - \frac{1}{ \tilde \kappa}{ D ^2 \tilde\sigma_2 } \right],
\end{align}
with $\tilde \kappa =8\pi \tilde G$ and $ \tilde G ={\hat G}/{\int dy^2} = {\hat G \mu_2}/{(2\pi)}$.
Taking a rescaling $\tilde A^2_{\tilde a}/\sqrt{2\tilde \kappa} \to \tilde A^2_{\tilde a}$ and integration by parts, we can further simplified this action to
\begin{align} \label{5D-effective-action-3}
S =\int d^5 x \sqrt{-\tilde g} e^{\tilde\sigma_2 } \left[ \frac{ \tilde R }{2 \tilde \kappa}  - \frac{1}{4} {f_2^2} e^{2{\tilde\sigma_2}} \tilde F^{2\tilde\alpha \tilde\beta} \tilde F^2_{\tilde\alpha \tilde \beta}  \right],
\end{align}
Very interestingly, thanks to a conformal transformation,
\begin{equation}
\tilde {\tilde g}_{\tilde a \tilde b} = e^{\tilde\sigma_2}  {\tilde g}_{\tilde a \tilde b},
\end{equation}
the action \eqref{5D-effective-action-3} becomes a familiar one such as
\begin{align} \label{5D-effective-action-4}
S =\int d^5 x \sqrt{-\tilde{\tilde g}} \left[ \frac{ \tilde {\tilde R} }{2 \tilde \kappa}  - \frac{1}{4 } {f_2^2} e^{3{\tilde\sigma_2}} \tilde {\tilde F}^{2 \tilde \alpha \tilde \beta} \tilde {\tilde F}^2_{\tilde \alpha \tilde \beta}  -\frac{3}{4 \tilde \kappa}   \left(\tilde \partial_{\tilde a} \tilde\sigma_2 \right) \left(\tilde \partial^{\tilde a} \tilde\sigma_2 \right) \right].
\end{align}
Now, we would like to discuss the issue related to the existence of the last two terms in the final action \eqref{5D-effective-action-4}. Remember that $\tilde\sigma_2$ and $\tilde A^2_{\tilde a}$ have been generally assumed to be dependent on not only the $4D$ coordinates $x^a$ but also the fifth one $y^1$. This fact will make the $S^1$ compactification of fifth dimension very complicated to complete in order to derive the corresponding effective $4D$ action. A resolution to this issue is perhaps to require that both of these field depend only the $4D$ coordinates, similar to the previous case, i.e.,
\begin{equation}
\tilde \sigma_2 = \tilde \sigma_2(x),\quad \tilde A^2_{\tilde a} = \tilde A^2_{\tilde a} (x).
\end{equation}
 If so, it seems that we would end up with an effective $4D$ action involving two scalar and two vector fields, similar to the previous case. However, the resulted effective $4D$ action will turn out to be different from that derived in the $T^2$ compactification approach. This is indeed a significant gap between the two approaches of dimensional reduction considered in this paper. 

Indeed, one can easily show that
\begin{equation}
\tilde {\tilde R} = R -\frac{1}{4}f_1^2 e^{2\sigma_1} F^{1\alpha\beta} F^1_{\alpha\beta}- 2 \left(\partial_{ a} \sigma_1 \right) \left(\partial^{ a } \sigma_1 \right) - 2 {D^2 \sigma_1},
\end{equation}
for the following $5D$ spacetime given by
\begin{equation}
ds^2= g_{\tilde a \tilde b} (x,y^1) dX^{\tilde a} dX^{\tilde b} = g_{ab} (x) dx^a dx^b +e^{2\sigma_1 (x)} \left[dy^1 -f_1 A_a^1 (x)dx^a \right]^2.
\end{equation}
For convenience, we will rewrite the last two terms in the action \eqref{5D-effective-action-4} as follows
\begin{align}
\frac{1}{4 } {f_2^2} e^{3{\tilde\sigma_2}} \tilde {\tilde F}^{2\tilde \alpha \tilde\beta} \tilde {\tilde F}^2_{\tilde \alpha \tilde \beta} = \frac{1}{4 } {f_2^2} e^{3{\sigma_2}}  { F}^{2\alpha \beta} {F}^2_{\alpha \beta},\quad \frac{3}{2 \tilde \kappa}   \left(\tilde \partial_{\tilde a} \tilde\sigma_2 \right) \left(\tilde \partial^{\tilde a} \tilde\sigma_2 \right) = \frac{3}{2 \kappa}   \left( \partial_{a} \sigma_2 \right) \left( \partial^{a} \sigma_2 \right),
\end{align}
where all ``tilde'' have been omitted on the right hand sides. As a result, the action \eqref{5D-effective-action-4} is reduced, after taking the second $S^1$ compactification for the fifth dimension $y^1$ and a suitable rescaling of the second vector field, to
\begin{equation} 
S = \int d^4x \sqrt{-g}e^{\sigma_1} \left[ \frac{R}{2\kappa} -\frac{1}{4}f_1^2 e^{2\sigma_1} F^{1\alpha\beta} F^1_{\alpha\beta} -\frac{1}{\kappa} \left(\partial_{ a} \sigma_1 \right) \left(\partial^{ a } \sigma_1 \right) -\frac{1}{\kappa} {D^2 \sigma_1} -\frac{1}{4 } {f_2^2} e^{3{\sigma_2}}  { F}^{2\alpha \beta} {F}^2_{\alpha \beta} - \frac{1}{2 \kappa}   \left( \partial_{a} \sigma_2 \right) \left( \partial^{a} \sigma_2 \right) \right].
\end{equation}
 Furthermore, this action can be further simplified, after taking integrations by part, as
\begin{equation} \label{Jordan-1}
S = \int d^4x \sqrt{-g}e^{\sigma_1} \left[ \frac{R}{2\kappa} -\frac{1}{4}f_1^2 e^{2\sigma_1} F^{1\alpha\beta} F^1_{\alpha\beta}-\frac{1}{4 } {f_2^2} e^{3{\sigma_2}}  { F}^{2\alpha \beta} {F}^2_{\alpha \beta}   - \frac{3}{4 \kappa}   \left( \partial_{a} \sigma_2 \right) \left( \partial^{a} \sigma_2 \right) \right].
\end{equation}
This action is completely different from that shown in Eq. \eqref{Jordan}. And this is the main result of our paper. For quick comparisons, main results of the two approaches of dimensional reduction will be summarized  in the Table \ref{tab:my-table}. 

Similar to the previous section, one can transform the above action written in the Jordan frame to that given in the Einstein frame,
 \begin{equation} \label{Einstein-1}
S = \int d^4x \sqrt{-\bar{\bar g}} \left[ \frac{\bar{\bar R}}{2\kappa} -\frac{1}{4}f_1^2 e^{3\sigma_1} \bar {\bar F}^{1\alpha\beta} \bar {\bar F}^1_{\alpha\beta}-\frac{1}{4 } {f_2^2} e^{\sigma_1+ 3{\sigma_2}}  \bar{\bar F}^{2\alpha \beta} \bar{\bar F}^2_{\alpha \beta}  -\frac{3}{4\kappa} \left(\bar{\bar \partial}_{ a} \sigma_1 \right) \left(\bar{\bar\partial}^{ a } \sigma_1 \right) - \frac{3}{4 \kappa}  \left(\bar{\bar \partial}_{a} \sigma_2 \right) \left( \bar{\bar\partial}^{a} \sigma_2 \right) \right],
\end{equation}
by using the following conformal transformation,
 \begin{equation}
 \bar{\bar g}_{ab}= e^{\sigma_1 } g_{ab},
 \end{equation}
 along with the related ones,
 \begin{align}
\sqrt{-g} &= e^{-2\sigma_1} \sqrt{-\bar{\bar g}},\\
R &= e^{\sigma_1} \left[ \bar{\bar R} -\frac{3}{2}( \bar{ \bar\partial}_a \sigma_1)(  \bar{\bar\partial}^a \sigma_1)  \right],\\
\sqrt{-g} F^2 &= \sqrt{-\bar{\bar g}} { \bar {\bar F}}^2,\\
(\partial_a \sigma_1) (\partial^a \sigma_1) &= e^{\sigma_1} (\bar {\bar \partial}_a \sigma_1)( \bar{\bar\partial}^a \sigma_1),\\
(\partial_a \sigma_2)( \partial^a \sigma_2) &= e^{\sigma_1} (\bar{\bar \partial}_a \sigma_2)( \bar{\bar\partial}^a \sigma_2).
\end{align}
So far, we have completed the derivation of 4D effective action due to the indirect $6D \to 5D\to 4D$ reduction. The significant gap between the direct and indirect spatial reductions considered in this paper is the existence of the mixed kinetic term of two scalar fields $\sigma_1$ and $\sigma_2$. 
		\begin{center}
	\begin{sidewaystable}[]
		\begin{tabular}{@{}|c|c|c|c|@{}}
			\hline
			\multirow{2}{*}{Objects} &
			\multirow{2}{*}{Direct $6D \to 4D$ reduction} &
			\multicolumn{2}{c|}{Indirect $6D\to 5D \to 4D$ reduction}\\
			 \cline{3-4}
			&{ }& {$6D\to 5D$ reduction }&
			{$5D\to 4D$ reduction } \\
			\hline
			\multirow{ 5}{*}{General coordinate indices}&{ }& { }&{ }\\
			& {$m,~ n =0,~1,~2,~3,~ 5,~6 $ }& {$m,~ n =0,~1,~2,~3,~ 5,~6 $ }& {$\tilde a,~ \tilde b =0,~1,~2,~3,~ 5$ } \\
			&{ }&{ }& { }\\
			&{ $a,~b=0,~ 1,~2,~3$}& {$\tilde a,~ \tilde b =0,~1,~2,~3,~ 5$ }& {$a,~b=0,~ 1,~2,~3$ }\\
			&{ }&{ }& { }\\
			\hline
			\multirow{ 5}{*}{Local coordinate indices}&{ }& { } &{ }\\
			& {$\mu,~\nu = 0,~1,~2,~3,~\bar 5,~\bar 6$ }& {$\mu,~\nu = 0,~1,~2,~3,~\bar 5,~\bar 6$}&{$\tilde \alpha,~\tilde \beta = 0,~1,~2,~3,~\bar 5$ } \\
			&{ }&{ }& { }\\
			& {$\alpha,~\beta,~\gamma,~\delta = 0,~1,~2,~3$} &{$\tilde\alpha,~\tilde\beta,~\tilde\gamma,~\tilde\delta = 0,~1,~2,~3,~\bar 5$ }& {$\alpha,~\beta,~\gamma,~\delta = 0,~1,~2,~3$ } \\
			&{ }&{ }&{ }\\
			\hline			
			\multirow{ 5}{*}{$6D$ (or $5D$) metric}&{ } &{ } &{ }\\
			&{ $ds^2= \hat g_{mn}(x,y^1,y^2)dX^mdX^n$} & {$ds^2 =\hat g_{mn}(x,y^1,y^2)dX^mdX^n $} & { $ds^2= g_{\tilde a \tilde b} (x,y^1) dX^{\tilde a} dX^{\tilde b}$}  \\
			&{ $ = g_{ab} (x) dx^a dx^b$ } & {$=\tilde g_{\tilde a \tilde b} (x,y^1) dX^{\tilde a} dX^{\tilde b}$} & { $= g_{ab} (x) dx^a dx^b$}  \\
			& {$+e^{2\sigma_1 (x)} \left[dy^1 -f_1 A_a^1 (x)dx^a \right]^2$}& {$ +e^{2\tilde  \sigma_2 (x,y^1)} \left[dy^2 -f_2 \tilde A_{\tilde a}^2 (x,y^1)dx^{\tilde a} \right]^2$} & { $+e^{2\sigma_1 (x)} \left[dy^1 -f_1 A_a^1 (x)dx^a \right]^2$} \\
			& {$+e^{2\sigma_2 (x)} \left[dy^2 -f_2 A_a^2 (x)dx^a \right]^2$} &{ } &{ } \\
			&{ }&{ }&{ }\\
			\hline
			\multirow{10}{*}{$6D$ (or $5D$) connection 1-form}&{ }&{ }&{ }\\
			&{ $\hat\omega^{\bar 5}{}_\alpha =  \left( \partial_a\sigma_1 \right) e_\alpha{}^a \hat\theta^{\bar 5} - \frac{1}{2} f_1 e^{\sigma_1} F^1_{\alpha\beta} \hat\theta^\beta,$} & { } & {$\tilde\omega^{\bar 5}{}_\alpha =  \left( \partial_a\sigma_1 \right) e_\alpha{}^a \tilde \theta^{\bar 5} - \frac{1}{2} f_1 e^{\sigma_1} F^1_{\alpha\beta} \tilde \theta^\beta,$ }  \\
			&{ }&{ } &{ }\\
			& {$\hat\omega^{\bar 5}{}_{\bar 5} = 0,$}& { }& {$\tilde \omega^{\bar 5}{}_{\bar 5} =0,$ } \\
			&{ }&{ }&{ }\\
			& {$\hat\omega^{\bar 6}{}_\alpha = \left( \partial_a\sigma_2 \right) e_\alpha{}^a \hat\theta^{\bar 6} - \frac{1}{2} f_2 e^{\sigma_2} F^2_{\alpha\beta} \hat\theta^\beta,$} &{ $ \hat\omega^{\bar 6}{}_{\tilde\alpha} = \left( \partial_{\tilde a } \tilde\sigma_2 \right) \hat e_{\tilde\alpha}{}^{\tilde a} \hat \theta^{\bar 6} -\frac{1}{2}f_2 e^{\tilde\sigma_2} \tilde F^2_{\tilde\alpha \tilde\beta} \hat\theta^{\tilde\beta},$ }& { } \\
			&{ }&{ } & { }\\
			& {$\hat\omega^{\bar 6}{}_{\bar 6} = 0,$} &{ $\hat\omega^{\bar 6}{}_{\bar 6} =0,$} &{ }\\
			&{ }&{ }& { }\\
			& {$\hat\omega^{\alpha}{}_{\beta} = \omega^\alpha{}_\beta + \frac{1}{2}f_1 e^{\sigma_1} F^{1\alpha}{}_\beta \hat\theta^{\bar 5}  +\frac{1}{2}f_2 e^{\sigma_2} F^{2\alpha}{}_\beta \hat\theta^{\bar 6}$} &{ $\hat\omega^{\tilde \alpha}{}_{\tilde \beta} = \tilde \omega^{\tilde \alpha}{}_{\tilde \beta} +\frac{1}{2}f_2 e^{\tilde\sigma_2} \tilde F^{2\tilde \alpha}{}_{\tilde \beta} \hat\theta^{\bar 6}$ } &{ $\tilde \omega^{\alpha}{}_{\beta} = \omega^\alpha{}_\beta + \frac{1}{2}f_1 e^{\sigma_1} F^{1\alpha}{}_\beta \tilde \theta^{\bar 5}$} \\
			&{ }&{ }&{ }\\
			\hline
			\multirow{ 9}{*}{$6D$ (or $5D$) Ricci scalar}&{ }& { } &{ }\\
			&{$\hat R = R - \frac{1}{4} {f_1^2} e^{2\sigma_1} F^{1\alpha \beta} F^1_{\alpha \beta}  - \frac{1}{4} {f_2^2} e^{2{\sigma_2}} F^{2\alpha \beta} F^2_{\alpha \beta}$ } &
			{$\hat R = \tilde R -\frac{1}{4}f_2^2 e^{2\tilde\sigma_2} \tilde F^{2\tilde\alpha\tilde\beta}\tilde F^2_{\tilde\alpha\tilde\beta}$} &{ $\tilde R = R -\frac{1}{4}f_1^2 e^{2\sigma_1} F^{1\alpha\beta} F^1_{\alpha\beta}$  }\\
			&{ }&{ }& { }\\
			& {$- 2  \left(\partial_a \sigma_1 \right) \left(\partial^a \sigma_1 \right) - 2  {D ^2 \sigma_1 } $ }& {$-2  \left(\partial_{\tilde a} \tilde\sigma_2 \right) \left(\partial^{\tilde a} \tilde\sigma_2 \right) - 2 {D^2 \tilde\sigma_2}$} &{$-2  \left(\partial_{ a} \sigma_1 \right) \left(\partial^{ a } \sigma_1 \right) -2 {D^2 \sigma_1}$ }\\
			&{ }&{ }& { }\\
			& {$-2   \left(\partial_a \sigma_2 \right) \left(\partial^a \sigma_2 \right) -2   { D^2 \sigma_2} $ }& { } &{ }\\
			&{ }&{ }&{ }\\
			& {$-2   \left(\partial_a \sigma_1 \right) \left(\partial^a\sigma_2 \right)$} &{ }&{ } \\
			&{ }&{ }&{ }\\
			\hline
			\multirow{ 3}{*}{Type of compactification}&{ }& { }&{ }\\	
			& {$T^2 = S^1 \times S^1$} &{ $S^1$}&{ $S^1$ } \\
			&{ }&{ }&{ }\\
			\hline
			\multirow{ 3}{*}{$4D$ effective action}&{ }& { }&{ }\\	
			 & {Displayed in Eq. \eqref{Jordan}} &{- } &{Displayed in Eq. \eqref{Jordan-1} }\\
			&{ }&{ }& { }\\
			\hline

			\multirow{ 3}{*}{Number of $4D$ matter fields}&{ }& { }& { }\\	
			 & {Two scalar and two vector fields} &{- }& { Two scalar and two vector fields } \\
			&{ }&{ }&{ }\\
			\hline
		\end{tabular}
		\caption{The summary of main results of the two approaches of dimensional reduction.}
		\label{tab:my-table}
	\end{sidewaystable}
\end{center}
\section{Conclusions} \label{final}
Motivated by the interesting works on the $5D$ Kaluza-Klein theory \cite{Thirring,Ichinose:2002kg}, we would like to propose in this paper a $6D$ extension of the Kaluza-Klein theory with two possible approaches of dimensional reduction, whose main results have been summarized in the Table \ref{tab:my-table}. The first one has been called the direct $6D \to 4D$ reduction, which is basically based on the $T^2 \equiv S^1 \times S^1$ compactification, while the second one has been called the indirect $6D \to 5D\to 4D$ reduction, which is possible via two separated $S^1$ compactifications. As a result,  the $4D$ effective action obtained via the second approach is different from that derived via the first approach. This is a significant gap between these two approaches.  In addition, it has shown in the second approach of dimensional reduction that the latter $S^1$ compactification cannot be possible if at least one field coming from the first $S^1$ compactification depends on the fifth coordinate. 
All these results  could therefore address an important question of which approach is more reliable than the other. In this case, observational and/or experimental data might play a judgment role. We will leave this issue to our future investigations. To end this section, we would like to mention that our present paper  contributes, in harmony with many seminal published works, e.g., see Refs. \cite{Freund:1980xh,pope,Randjbar-Daemi:1982opc,Denef:2007pq,Cvetic:2000dm}, one more ``natural'' explanation of the origin of multi vector and scalar field models to the current literature. Indeed, one could conjecture that any $4D$ gravity action involving $N$ vector fields and $N$ scalar fields could be originated from a $(4+N)-D$ pure gravity action using suitable KK compactifications. Additionally, the similar issue of the gap between different compactifications could be found in all higher-than-six-dimensional KK theories.
\begin{acknowledgments}
We would like to thank Prof. Min-Seok Seo very much for his useful suggestion, which helps us to improve our calculations. This study is partially  funded by Vietnam National Foundation for Science and Technology Development (NAFOSTED) under grant number 103.01-2023.50 (T.Q.D.). 
\end{acknowledgments}

\appendix 
\section{Non-vanishing components of $6D$ Riemann tensor} \label{app1}
In this Appendix, we derive explicitly the corresponding non-vanishing components of $6D$ Riemann tensor using the second Cartan's structure equation \eqref{second}. For $\mu=\alpha$ and $\nu=\beta$, we have
\begin{align}
&d\hat\omega_{\alpha\beta} +\hat\omega_{\alpha \sigma} \wedge \hat\omega^\sigma{}_\beta =\frac{1}{2} \hat R_{\alpha\beta\sigma \tau} \hat\theta^\sigma \wedge \hat\theta^\tau,
\end{align}
which can be fully expanded to be
\begin{align}
&d \omega_{\alpha \beta}  + \omega_{\alpha\gamma} \wedge \omega^\gamma{}_\beta -\frac{f_1^2}{2} \hat\eta_{\alpha\gamma} \partial_b \left(e^{2\sigma_1} F^{1\gamma}{}_\beta A^1_a\right) dX^b \wedge dX^a + \frac{f_1}{2} \hat\eta_{\alpha\gamma} \partial_b \left( e^{2\sigma_1} F^{1\gamma}{}_\beta \right) dX^b \wedge dX^5 \nonumber\\
&-\frac{f_2^2}{2} \hat\eta_{\alpha\gamma} \partial_b \left(e^{2\sigma_2} F^{2\gamma}{}_\beta A^2_a\right) dX^b \wedge dX^a + \frac{f_2}{2} \hat\eta_{\alpha\gamma} \partial_b \left( e^{2\sigma_2} F^{2\gamma}{}_\beta \right) dX^b \wedge dX^6  \nonumber\\
&+\frac{f_1 f_2}{4} e^{\sigma_1+\sigma_2} \left( F^1_{\alpha \gamma} F^{2\gamma}{}_\beta -F^{2}_{\alpha\gamma} F^{1\gamma}{}_\beta \right) \hat\theta^{\bar 5} \wedge \hat\theta^{\bar 6} \nonumber\\
&+\frac{f_1}{4}\hat\eta_{\alpha\delta}  \hat\eta^{\delta \gamma} {\hat \eta_{\bar 5 \bar 5}} e^{\sigma_1}  \left(\partial_a \sigma_1\right) e_\gamma{}^a F^1_{\beta\epsilon'} \hat\theta^{\bar 5} \wedge \hat\theta^{\epsilon'} +\frac{f_1}{4}\hat\eta_{\alpha\delta}  \hat\eta^{\delta \gamma} {\hat \eta_{\bar 5 \bar 5}} e^{\sigma_1} \left(\partial_b \sigma_1\right) e_\beta{}^b  F^1_{\gamma\epsilon} \hat\theta^{\epsilon} \wedge \hat\theta^{\bar 5} \nonumber\\
& -\frac{f_1^2}{4}\hat\eta_{\alpha\delta}  \hat\eta^{\delta \gamma} {\hat \eta_{\bar 5 \bar 5}}  e^{2\sigma_1} F^1_{\gamma \epsilon} F^1_{\beta\epsilon'} \hat\theta^\epsilon \wedge \hat\theta^{\epsilon'} \nonumber\\
&+\frac{f_2}{4}\hat\eta_{\alpha\delta}  \hat\eta^{\delta \gamma} {\hat \eta_{\bar 6 \bar 6}} e^{\sigma_2} \left(\partial_a \sigma_2\right) e_\gamma{}^a  F^2_{\beta\epsilon'} \hat\theta^{\bar 6} \wedge \hat\theta^{\epsilon'} +\frac{f_2}{4}\hat\eta_{\alpha\delta}  \hat\eta^{\delta \gamma} {\hat \eta_{\bar 6 \bar 6}} e^{\sigma_2}  \left(\partial_b \sigma_2\right) e_\beta{}^b F^2_{\gamma\epsilon} \hat\theta^{\epsilon} \wedge \hat\theta^{\bar 6} \nonumber\\
& -\frac{f_2^2}{4}\hat\eta_{\alpha\delta}  \hat\eta^{\delta \gamma} {\hat \eta_{\bar 6 \bar 6}} e^{2\sigma_2} F^2_{\gamma \epsilon} F^2_{\beta\epsilon'} \hat\theta^\epsilon \wedge \hat\theta^{\epsilon'} \nonumber\\
=&~ \frac{1}{2} \left(\hat R_{\alpha\beta\gamma \delta} \hat\theta^\gamma \wedge \hat\theta^\delta +\hat R_{\alpha\beta \bar 5 \delta} \hat\theta^{\bar 5} \wedge \hat\theta^\delta +\hat R_{\alpha\beta \delta \bar 5}  \hat\theta^\delta \wedge \hat\theta^{\bar 5} +\hat R_{\alpha\beta \bar 6 \delta} \hat\theta^{\bar 6} \wedge \hat\theta^\delta +\hat R_{\alpha\beta \delta \bar 6}  \hat\theta^\delta \wedge \hat\theta^{\bar 6}  +\hat R_{\alpha\beta \bar 5 \bar 6} \hat\theta^{\bar 5} \wedge \hat\theta^{\bar 6} +\hat R_{\alpha\beta \bar 6 \bar 5}  \hat\theta^{\bar 6} \wedge \hat\theta^{\bar 5} \right) .
\end{align}
Now, thanks to the relations shown in Eqs. \eqref{hat-theta-1}-\eqref{hat-e-3} we are able to have
\begin{align}
&dX^a = e_\alpha{}^a \hat\theta^\alpha,\nonumber\\
 &dX^5=e^{-\sigma_1} \hat\theta^{\bar 5} +f_1A^1_a dX^a, \nonumber\\
& dX^6=e^{-\sigma_2} \hat\theta^{\bar 6} +f_2A^2_a dX^a.
 \end{align}
 Additionally, it turns out that the $4D$ version of second Cartan's equation is given by
 \begin{equation}
 d \omega_{\alpha \beta}  + \omega_{\alpha\gamma} \wedge \omega^\gamma{}_\beta = \frac{1}{2}R_{\alpha\beta\gamma\delta} \theta^\gamma \wedge \theta^\delta,
 \end{equation}
 Armed with these useful results, we are able to simplify the above lengthy equation as
 \begin{align}
& \frac{1}{2}R_{\alpha\beta\gamma\delta} \hat\theta^\gamma \wedge \hat\theta^\delta  -\frac{f_1^2}{4} e^{2\sigma_1} F^1_{\alpha\beta} F^1_{\gamma\delta}  \hat\theta^\gamma \wedge \hat\theta^\delta +\frac{f_1}{2} \hat\eta_{\alpha \gamma} e^{-\sigma_1} \partial_b \left(e^{2\sigma_1} F^{1\gamma}{}_\beta \right) e_\delta{}^b \hat\theta^\delta \wedge \hat\theta^{\bar 5} \nonumber\\
 &-\frac{f_2^2}{4} e^{2\sigma_2} F^2_{\alpha\beta} F^2_{\gamma\delta}  \hat\theta^\gamma \wedge \hat\theta^\delta +\frac{f_2}{2} \hat\eta_{\alpha \gamma} e^{-\sigma_2} \partial_b \left(e^{2\sigma_2} F^{2\gamma}{}_\beta \right) e_\delta{}^b \hat\theta^\delta \wedge \hat\theta^{\bar 6} \nonumber\\
 & +\frac{f_1 f_2}{4} e^{\sigma_1+\sigma_2} \left( F^1_{\alpha \gamma} F^{2\gamma}{}_\beta -F^{2}_{\alpha\gamma} F^{1\gamma}{}_\beta \right) \hat\theta^{\bar 5} \wedge \hat\theta^{\bar 6} \nonumber\\
 &+\frac{f_1}{4} e^{\sigma_1}  \left(\partial_a \sigma_1\right)  e_\alpha{}^a F^1_{\beta\epsilon'} \hat\theta^{\bar 5} \wedge \hat\theta^{\epsilon'} + \frac{f_1}{4} e^{\sigma_1}  \left(\partial_b \sigma_1\right) e_\beta{}^b F^1_{\alpha\epsilon} \hat\theta^{\epsilon} \wedge \hat\theta^{\bar 5} - \frac{f_1^2}{4} e^{2\sigma_1} F^1_{\alpha \epsilon} F^1_{\beta\epsilon'} \hat\theta^\epsilon \wedge \hat\theta^{\epsilon'} \nonumber\\
&+\frac{f_2}{4} e^{\sigma_2}  \left(\partial_a \sigma_2\right)  e_\alpha{}^a F^2_{\beta\epsilon'} \hat\theta^{\bar 6} \wedge \hat\theta^{\epsilon'} + \frac{f_2}{4} e^{\sigma_2}  \left(\partial_b \sigma_2\right) e_\beta{}^b F^2_{\alpha\epsilon} \hat\theta^{\epsilon} \wedge \hat\theta^{\bar 6} -\frac{f_2^2}{4} e^{2\sigma_2} F^2_{\alpha \epsilon} F^2_{\beta\epsilon'} \hat\theta^\epsilon \wedge \hat\theta^{\epsilon'} \nonumber\\
=&~\frac{1}{2} \left(\hat R_{\alpha\beta\gamma \delta} \hat\theta^\gamma \wedge \hat\theta^\delta +\hat R_{\alpha\beta \bar 5 \delta} \hat\theta^{\bar 5} \wedge \hat\theta^\delta +\hat R_{\alpha\beta \delta \bar 5}  \hat\theta^\delta \wedge \hat\theta^{\bar 5} +\hat R_{\alpha\beta \bar 6 \delta} \hat\theta^{\bar 6} \wedge \hat\theta^\delta +\hat R_{\alpha\beta \delta \bar 6}  \hat\theta^\delta \wedge \hat\theta^{\bar 6}  +\hat R_{\alpha\beta \bar 5 \bar 6} \hat\theta^{\bar 5} \wedge \hat\theta^{\bar 6} +\hat R_{\alpha\beta \bar 6 \bar 5}  \hat\theta^{\bar 6} \wedge \hat\theta^{\bar 5} \right) .
 \end{align}
 Rearranging the terms in the left hand side in order to be compatible with the ones in the right hand side leads to
 \begin{align}
 & \frac{1}{2}R_{\alpha\beta\gamma\delta} \hat\theta^\gamma \wedge \hat\theta^\delta -\frac{f_1^2}{4} e^{2\sigma_1} F^1_{\alpha\beta} F^1_{\gamma\delta}  \hat\theta^\gamma  \wedge \hat\theta^\delta - \frac{f_2^2}{4} e^{2\sigma_2} F^2_{\alpha\beta} F^2_{\gamma\delta}  \hat\theta^\gamma \wedge \hat\theta^\delta \nonumber\\
 & +\frac{f_1^2}{8}e^{2\sigma_1} \left(F^1_{\alpha \delta}F^1_{\beta \gamma} -F^1_{\alpha \gamma}F^1_{\beta \delta} \right)  \hat\theta^\gamma \wedge \hat\theta^\delta +\frac{f_2^2}{8}e^{2\sigma_2} \left(F^2_{\alpha \delta}F^2_{\beta \gamma} -F^2_{\alpha \gamma}F^2_{\beta \delta} \right)  \hat\theta^\gamma \wedge \hat\theta^\delta =  \frac{1}{2} \hat R_{\alpha\beta\gamma \delta} \hat\theta^\gamma \wedge \hat\theta^\delta, \\
 &\frac{f_1}{2}  e^{-\sigma_1} \partial_b \left(e^{2\sigma_1} F^1_{\alpha\beta} \right) e_\delta{}^b \hat\theta^\delta \wedge \hat\theta^{\bar 5} - \frac{f_1}{4} e^{\sigma_1}  \partial_a \sigma_1 \left( e_\alpha{}^a F^1_{\beta\delta}  - e_\beta{}^a F^1_{\alpha\delta} \right) \hat\theta^{\delta} \wedge \hat\theta^{\bar 5} = \hat R_{\alpha\beta \delta \bar 5}  \hat\theta^\delta \wedge \hat\theta^{\bar 5}, \\
 &\frac{f_2}{2}  e^{-\sigma_2} \partial_b \left(e^{2\sigma_2} F^2_{\alpha\beta} \right) e_\delta{}^b \hat\theta^\delta \wedge \hat\theta^{\bar 6} - \frac{f_2}{4} e^{\sigma_2}  \partial_a \sigma_2 \left( e_\alpha{}^a F^2_{\beta\delta}  - e_\beta{}^a F^2_{\alpha\delta} \right) \hat\theta^{\delta} \wedge \hat\theta^{\bar 6} = \hat R_{\alpha\beta \delta \bar 6}  \hat\theta^\delta \wedge \hat\theta^{\bar 6}, \\
 &\frac{f_1 f_2}{4} e^{\sigma_1+\sigma_2} \left( F^1_{\alpha \gamma} F^{2\gamma}{}_\beta -F^{2}_{\alpha\gamma} F^{1\gamma}{}_\beta \right) \hat\theta^{\bar 5} \wedge \hat\theta^{\bar 6} =\hat R_{\alpha\beta \bar 5 \bar 6} \hat\theta^{\bar 5} \wedge \hat\theta^{\bar 6}.
 \end{align} 
 It is noted that we have used the following result,
 \begin{align}
 - \frac{f_1^2}{4} e^{2\sigma_1} F^1_{\alpha \epsilon} F^1_{\beta\epsilon'} \hat\theta^\epsilon \wedge \hat\theta^{\epsilon'}  = - \frac{f_1^2}{8}e^{2\sigma_1} F^1_{\alpha \epsilon} F^1_{\beta\epsilon'} \hat\theta^\epsilon \wedge \hat\theta^{\epsilon'} -  \frac{f_1^2}{8}e^{2\sigma_1}  F^1_{\alpha \epsilon'} F^1_{\beta\epsilon} \hat\theta^{\epsilon'} \wedge \hat\theta^{\epsilon}.
 \end{align}
 Hence, we have the following results,
 \begin{align}
 \hat R_{\alpha\beta\gamma \delta}  =&~ R_{\alpha\beta\gamma\delta} -\frac{f_1^2}{2} e^{2\sigma_1} F^1_{\alpha\beta} F^1_{\gamma\delta}- \frac{f_2^2}{2} e^{2\sigma_2} F^2_{\alpha\beta} F^2_{\gamma\delta} \nonumber\\
 &+\frac{f_1^2}{4}e^{2\sigma_1} \left(F^1_{\alpha \delta}F^1_{\beta \gamma} -F^1_{\alpha \gamma}F^1_{\beta \delta} \right)+\frac{f_2^2}{4}e^{2\sigma_2} \left(F^2_{\alpha \delta}F^2_{\beta \gamma} -F^2_{\alpha \gamma}F^2_{\beta \delta} \right).
 \end{align}
For $\mu=\bar 5$ and $\nu=\bar 5$ as well as  $\mu=\bar 6$ and $\nu=\bar 6$, we have
\begin{equation}
\hat R_{\bar 5 \bar 5 \sigma \tau}  =0; ~\hat R_{\bar 6 \bar 6 \sigma \tau}  =0,
\end{equation}
as expected.
For $\mu=\bar 5$ and $\nu=\bar 6$, we have
\begin{align}
d\hat\omega_{\bar 5\bar 6} +\hat\omega_{\bar 5 \sigma} \wedge \hat\omega^\sigma{}_{\bar 6} =\frac{1}{2} \hat R_{\bar 5 \bar 6 \sigma \tau} \hat\theta^\sigma \wedge \hat\theta^\tau,
\end{align}
which can be fully expanded to be
\begin{align}
 & - \hat\eta_{\bar 5 \bar 5} \hat\eta_{\bar 6\bar 6}  \hat\eta^{\alpha\beta}  \left[\left(\partial_a \sigma_1\right) \left( \partial_b \sigma_2 \right)  e_\alpha{}^a  e_\beta{}^b  \hat\theta^{\bar 5} \wedge \hat\theta^{\bar 6} - \frac{1}{2} f_1 e^{\sigma_1} F^1_{\alpha\gamma} \left(\partial_b \sigma_2 \right) e_\beta{}^b   \hat\theta^\gamma \wedge \hat\theta^{\bar 6} \right. \nonumber\\
&\left. - \frac{1}{2} f_2 \left(\partial_a \sigma_1 \right) e_\alpha{}^a e^{\sigma_2} F^2_{\beta\delta}  \hat\theta^{\bar 5} \wedge \hat\theta^\delta + \frac{1}{4} f_1f_2 e^{\sigma_1} F^1_{\alpha\gamma} e^{\sigma_2} F^2_{\beta\delta} \hat\theta^\gamma \wedge  \hat\theta^\delta \right] = \frac{1}{2} \hat R_{\bar 5 \bar 6 \alpha\beta} \hat\theta^\alpha \wedge \hat\theta^\beta \nonumber\\
& + \frac{1}{2} \hat R_{\bar 5 \bar 6 \bar 5 \bar 6} \hat\theta^{\bar 5} \wedge \hat\theta^{\bar 6}+\frac{1}{2} \hat R_{\bar 5 \bar 6 \bar 6 \bar 5} \hat\theta^{\bar 6} \wedge \hat\theta^{\bar 5}  +\frac{1}{2} \hat R_{\bar 5 \bar 6 \bar 5 \alpha} \hat\theta^{\bar 5} \wedge \hat\theta^\alpha +\frac{1}{2} \hat R_{\bar 5 \bar 6 \alpha  \bar 5} \hat\theta^\alpha \wedge  \hat\theta^{\bar 5} +\frac{1}{2} \hat R_{\bar 5 \bar 6 \bar 6 \alpha} \hat\theta^{\bar 6} \wedge \hat\theta^\alpha  +\frac{1}{2} \hat R_{\bar 5 \bar 6 \alpha \bar 6 } \hat\theta^\alpha \wedge \hat\theta^{\bar 6}  .
\end{align}
As a result, we have, according to this equation, that
\begin{align}
\hat R_{\bar 5 \bar 6 \bar 5 \bar 6}  &= -\hat R_{\bar 5 \bar 6 \bar 6 \bar 5}= - g^{ab}  \left( \partial_a \sigma_1 \right) \left( \partial_b \sigma_2 \right) ,
\end{align}
where $g^{ab} = e_\alpha{}^a e_\beta{}^b \hat\eta^{\alpha\beta}$. 
For $\mu=\bar 6$ and $\nu=\bar 5$, we have
\begin{align}
&d\hat\omega_{\bar 6\bar 5} +\hat\omega_{\bar 6 \sigma} \wedge \hat\omega^\sigma{}_{\bar 5} =\frac{1}{2} \hat R_{\bar 6 \bar 5 \sigma \tau} \hat\theta^\sigma \wedge \hat\theta^\tau,
\end{align}
which becomes as
\begin{equation}
-\hat\eta_{\bar 5 \bar 5} \hat\eta_{\bar 6\bar 6}  \hat\eta^{\alpha\beta} \left[\left(\partial_a \sigma_2 \right) e_\alpha{}^a \hat\theta^{\bar 6} -\frac{f_2}{2} e^{\sigma_2} F^2_{\alpha\gamma} \hat\theta^\gamma \right] \wedge \left[ \left(\partial_b \sigma_1 \right) e_\beta{}^b \hat\theta^{\bar 5} -\frac{f_1}{2} e^{\sigma_1} F^1_{\beta\delta} \hat\theta^\delta \right]=\frac{1}{2} \hat R_{\bar 6 \bar 5 \sigma \tau} \hat\theta^\sigma \wedge \hat\theta^\tau.
\end{equation}
As a result, this equation implies that
\begin{align}
\hat R_{\bar 6 \bar 5 \bar 5 \bar 6}  &= -\hat R_{\bar 6 \bar 5 \bar 6 \bar 5}= g^{ab} \left( \partial_a \sigma_2 \right) \left( \partial_b \sigma_1 \right),
\end{align}
as expected. For $\mu=\bar 5$ and $\nu=\alpha$, we have
\begin{align}
d\hat\omega_{\bar 5\alpha} +\hat\omega_{\bar 5 \sigma} \wedge \hat\omega^\sigma{}_{\alpha} =\frac{1}{2} \hat R_{\bar 5 \alpha \sigma \tau} \hat\theta^\sigma \wedge \hat\theta^\tau,
\end{align}
which implies that
\begin{align}
& \hat\eta_{\bar 5 \bar 5} \partial_b \left[  \left( \partial_a\sigma_1 \right) e_\alpha{}^a e^{\sigma_1} \right] e_\beta{}^b \hat\theta^\beta \wedge \left(e^{-\sigma_1}\hat\theta^{\bar 5} +f_1 A^1_c e_\gamma{}^c \hat\theta^\gamma \right) \nonumber\\
& - f_1 \hat\eta_{\bar 5 \bar 5} \partial_c \left[  \left( \partial_a\sigma_1 \right) e_\alpha{}^a A^1_b e^{\sigma_1}  \right] e_\beta{}^b e_\gamma{}^c \hat\theta^\gamma \wedge  \hat\theta^\beta \nonumber\\
&- \frac{1}{2}f_1\hat\eta_{\bar 5 \bar 5} \partial_c \left(e^{\sigma_1} F^1_{\alpha\beta} e^\beta{}_b \right) e_\delta{}^b e_\gamma{}^c   \hat\theta^\gamma \wedge \hat\theta^\delta + \hat\eta_{\bar 5\bar 5} \left( \partial_b\sigma_1 \right) e_\beta{}^b\left(\partial_c e^\beta{}_d \right) e_\alpha{}^c e_\gamma{}^d \hat\theta^{\bar 5} \wedge \hat\theta^\gamma \nonumber\\
&- \frac{1}{2} \hat\eta_{\bar 5\bar 5}  f_1 e^{\sigma_1} F^1_{\beta\gamma} \left(\partial_c e^\beta{}_b \right) e_\alpha{}^c e_\delta{}^b \hat\theta^\gamma \wedge \hat\theta^\delta +\frac{1}{2} \hat\eta_{\bar 5\bar 5} {f_2} e^{\sigma_2}  \left( \partial_b\sigma_1 \right) e_\beta{}^b  F^{2\beta}{}_\alpha \hat\theta^{\bar 5} \wedge \hat\theta^{\bar 6} \nonumber\\
&- \frac{1}{4}\hat\eta_{\bar 5\bar 5} f_1^2 e^{2 \sigma_1} F^1_{\beta\gamma} F^{1\beta}{}_\alpha \hat\theta^{\gamma} \wedge \hat\theta^{\bar 5} - \frac{1}{4}\hat\eta_{\bar 5\bar 5} f_1f_2 e^{\sigma_1+\sigma_2} F^1_{\beta\gamma} F^{2\beta}{}_\alpha \hat\theta^{\gamma} \wedge \hat\theta^{\bar 6} \nonumber\\
=& ~\frac{1}{2} \hat R_{\bar 5 \alpha \beta \gamma} \hat\theta^\beta \wedge \hat\theta^\gamma +\hat R_{\bar 5 \alpha \bar 5 \beta} \hat\theta^{\bar 5} \wedge \hat\theta^\beta+\hat R_{\bar 5 \alpha \bar 6 \beta} \hat\theta^{\bar 6} \wedge \hat\theta^\beta +\hat R_{\bar 5 \alpha \bar 5 \bar 6} \hat\theta^{\bar 5} \wedge \hat\theta^{\bar 6},
\end{align}
here we have used the result, 
\begin{equation}
\omega^\alpha{}_\beta =\left(\partial_b e^\alpha{}_a \right) e_\beta{}^b dX^a =\left(\partial_b e^\alpha{}_a \right) e_\beta{}^b e_\gamma{}^a\hat\theta^\gamma,
\end{equation}
which is derived from the first $4D$ Cartan's structure equation, $d\theta^\alpha+\omega^\alpha{}_\beta \wedge \theta^\beta=0$. Grouping the terms having the same $\hat\theta \wedge \hat\theta$ leads us to the following results,
\begin{align}
\hat R_{\bar 5 \alpha \bar 5 \beta} =& - \left[\partial_b \left(\partial_a \sigma_1 \right) \right] e_\alpha{}^a e_\beta{}^b - \left(\partial_a \sigma_1 \right)\left(\partial_b \sigma_1 \right) e_\alpha{}^a e_\beta{}^b - \left(\partial_a \sigma_1 \right)\left(\partial_b e_\alpha{}^a \right) e_\beta{}^b \nonumber\\
&  - \left(\partial_a \sigma_1 \right)\left(\partial_b e_\beta{}^a \right) e_\alpha{}^b  +\frac{1}{4} f_1^2 e^{2 \sigma_1} F^1_{\gamma\beta} F^{1\gamma}{}_\alpha.
\end{align}
For $\mu=\alpha$ and $\nu=\bar 5$, we have
\begin{align}
&d\hat\omega_{\alpha\bar 5} +\hat\omega_{\alpha \sigma} \wedge \hat\omega^\sigma{}_{\bar 5} =\frac{1}{2} \hat R_{\alpha \bar 5 \sigma \tau} \hat\theta^\sigma \wedge \hat\theta^\tau,
\end{align}
which can be shown to admit the following results as
\begin{align}
\hat R_{\alpha \bar 5  \bar 5 \beta} = - \hat R_{\bar 5 \alpha \bar 5 \beta}.
\end{align}
Similar to the case of $\mu=\bar 5$ and $\nu=\alpha$, we have for the case of $\mu=\bar 6$ and $\nu=\alpha$,
\begin{align}
&d\hat\omega_{\bar 6 \alpha} +\hat\omega_{\bar 6 \sigma} \wedge \hat\omega^\sigma{}_{\alpha} =\frac{1}{2} \hat R_{\bar 6 \alpha \sigma \tau} \hat\theta^\sigma \wedge \hat\theta^\tau,
\end{align}
the following results,
\begin{align}
\hat R_{\bar 6 \alpha \bar 6 \beta} =& - \left[\partial_b \left(\partial_a \sigma_2 \right) \right] e_\alpha{}^a e_\beta{}^b - \left(\partial_a \sigma_2 \right)\left(\partial_b \sigma_2 \right) e_\alpha{}^a e_\beta{}^b - \left(\partial_a \sigma_2 \right)\left(\partial_b e_\alpha{}^a \right) e_\beta{}^b\nonumber\\
&- \left(\partial_a \sigma_2 \right)\left(\partial_b e_\beta{}^a \right) e_\alpha{}^b  +\frac{1}{4} f_2^2 e^{2 \sigma_2} F^2_{\gamma\beta} F^{2\gamma}{}_\alpha.
\end{align}
It is straightforward to have the fact that, $\hat R_{\alpha \bar 6  \bar 6 \beta} = - \hat R_{\bar 6 \alpha \bar 6 \beta}$, for $\mu=\bar 6$ and $\nu =\alpha$. 
\section{Non-vanishing components of $6D$ Ricci tensor} \label{app2}
Given all non-vanishing relevant  components of $6D$ Riemann tensor derived in the previous Appendix, we are now able to define the corresponding non-vanishing components of $6D$ Ricci tensor, $\hat R_{\mu\nu} = \hat R^\rho{}_{\mu\rho\nu} \equiv \hat \eta^{\kappa \rho} \hat R_{\kappa \mu \rho \nu}$, as follows
\begin{align}
\hat R_{\alpha\beta} =& - \hat R_{0 \alpha 0 \beta} +\hat R_{1 \alpha 1 \beta} +\hat R_{2 \alpha 2 \beta} +\hat R_{3 \alpha 3 \beta} +\hat R_{\bar 5 \alpha \bar 5 \beta} + \hat R_{\bar 6 \alpha \bar 6 \beta} \nonumber\\
=& -R_{0\alpha 0\beta} +\frac{f_1^2}{2}e^{2\sigma_1} F^1_{0\alpha}F^1_{0\beta}+\frac{f_2^2}{2}e^{2\sigma_2} F^2_{0\alpha}F^2_{0\beta} \nonumber\\
&- \frac{f_1^2}{4}e^{2\sigma_1} \left(F^1_{0\beta}F^1_{\alpha 0} -F^1_{00}F^1_{\alpha \beta} \right)-\frac{f_2^2}{4}e^{2\sigma_2} \left(F^2_{0\beta}F^2_{\alpha 0} -F^2_{00}F^2_{\alpha \beta} \right) \nonumber\\
&+R_{1\alpha 1\beta} -\frac{f_1^2}{2}e^{2\sigma_1} F^1_{1\alpha}F^1_{1\beta}-\frac{f_2^2}{2}e^{2\sigma_2} F^2_{1\alpha}F^2_{1\beta} \nonumber\\
&+ \frac{f_1^2}{4}e^{2\sigma_1} \left(F^1_{1\beta}F^1_{\alpha 1} -F^1_{11}F^1_{\alpha \beta} \right)+\frac{f_2^2}{4}e^{2\sigma_2} \left(F^2_{1\beta}F^2_{\alpha 1} -F^2_{11}F^2_{\alpha \beta} \right) \nonumber\\
&+R_{2\alpha 2\beta} -\frac{f_1^2}{2}e^{2\sigma_1} F^1_{2\alpha}F^1_{2\beta}-\frac{f_2^2}{2}e^{2\sigma_2} F^2_{2\alpha}F^2_{2\beta} \nonumber\\
&+ \frac{f_1^2}{4}e^{2\sigma_1} \left(F^1_{2\beta}F^1_{\alpha 2} -F^1_{22}F^1_{\alpha \beta} \right)+\frac{f_2^2}{4}e^{2\sigma_2} \left(F^2_{2\beta}F^2_{\alpha 2} -F^2_{22}F^2_{\alpha \beta} \right) \nonumber\\
&+R_{3\alpha 3\beta} -\frac{f_1^2}{2}e^{2\sigma_1} F^1_{3\alpha}F^1_{3\beta}-\frac{f_2^2}{2}e^{2\sigma_2} F^2_{3\alpha}F^2_{3\beta} \nonumber\\
&+ \frac{f_1^2}{4}e^{2\sigma_1} \left(F^1_{3\beta}F^1_{\alpha 3} -F^1_{33}F^1_{\alpha \beta} \right)+\frac{f_2^2}{4}e^{2\sigma_2} \left(F^2_{3\beta}F^2_{\alpha 3} -F^2_{33}F^2_{\alpha \beta} \right) \nonumber\\
&-  \left[\partial_b \left(\partial_a \sigma_1 \right) \right] e_\alpha{}^a e_\beta{}^b - \left(\partial_a \sigma_1 \right)\left(\partial_b \sigma_1 \right) e_\alpha{}^a e_\beta{}^b -  \left(\partial_a \sigma_1 \right)\left(\partial_b e_\alpha{}^a \right) e_\beta{}^b  \nonumber\\
&-  \left(\partial_a \sigma_1 \right)\left(\partial_b e_\beta{}^a \right) e_\alpha{}^b +\frac{1}{4} f_1^2 e^{2 \sigma_1} F^1_{\gamma\beta} F^{1\gamma}{}_\alpha \nonumber\\
& - \left[\partial_b \left(\partial_a \sigma_2 \right) \right] e_\alpha{}^a e_\beta{}^b - \left(\partial_a \sigma_2 \right)\left(\partial_b \sigma_2 \right) e_\alpha{}^a e_\beta{}^b -  \left(\partial_a \sigma_2 \right)\left(\partial_b e_\alpha{}^a \right) e_\beta{}^b\nonumber\\
& -  \left(\partial_a \sigma_2 \right)\left(\partial_b e_\beta{}^a \right) e_\alpha{}^b +\frac{1}{4} f_2^2 e^{2 \sigma_2} F^2_{\gamma\beta} F^{2\gamma}{}_\alpha, 
\end{align}
\begin{align}
 \hat R_{\bar 5 \bar 5}=& -\hat R_{0\bar 5 0\bar 5} +\hat R_{1\bar 5 1\bar 5}+\hat R_{2\bar 5 2\bar 5}+\hat R_{3\bar 5 3\bar 5}+\hat R_{\bar 5 \bar 5 \bar 5 \bar 5} + \hat R_{\bar 6 \bar 5 \bar 6 \bar 5} \nonumber\\
 =&~ \left[\partial_b \left(\partial_a \sigma_1 \right) \right] e_0{}^a e_0{}^b +  \left(\partial_a \sigma_1 \right)\left(\partial_b \sigma_1 \right) e_0{}^a e_0{}^b +  \left(\partial_a \sigma_1 \right)\left(\partial_b e_0{}^a \right) e_0{}^b +  \left(\partial_a \sigma_1 \right)\left(\partial_b e_0{}^a \right) e_0{}^b \nonumber\\
 &  -\frac{1}{4} f_1^2 e^{2 \sigma_1} F^1_{\gamma0} F^{1\gamma}{}_0   -  \left[\partial_b \left(\partial_a \sigma_1 \right) \right] e_1{}^a e_1{}^b -  \left(\partial_a \sigma_1 \right)\left(\partial_b \sigma_1 \right) e_1{}^a e_1{}^b -  \left(\partial_a \sigma_1 \right)\left(\partial_b e_1{}^a \right) e_1{}^b \nonumber\\
& -  \left(\partial_a \sigma_1 \right)\left(\partial_b e_1{}^a \right) e_1{}^b  +\frac{1}{4} f_1^2 e^{2 \sigma_1} F^1_{\gamma1} F^{1\gamma}{}_1 -  \left[\partial_b \left(\partial_a \sigma_1 \right) \right] e_2{}^a e_2{}^b -  \left(\partial_a \sigma_1 \right)\left(\partial_b \sigma_1 \right) e_2{}^a e_2{}^b 
 \nonumber\\
 & -  \left(\partial_a \sigma_1 \right)\left(\partial_b e_2{}^a \right) e_2{}^b -  \left(\partial_a \sigma_1 \right)\left(\partial_b e_2{}^a \right) e_2{}^b +\frac{1}{4} f_1^2 e^{2 \sigma_1} F^1_{\gamma2} F^{1\gamma}{}_2  -  \left[\partial_b \left(\partial_a \sigma_1 \right) \right] e_3{}^a e_3{}^b  \nonumber\\
  & -  \left(\partial_a \sigma_1 \right)\left(\partial_b \sigma_1 \right) e_3{}^a e_3{}^b - \left(\partial_a \sigma_1 \right)\left(\partial_b e_3{}^a \right) e_3{}^b - \left(\partial_a \sigma_1 \right)\left(\partial_b e_3{}^a \right) e_3{}^b  +\frac{1}{4} f_1^2 e^{2 \sigma_1} F^1_{\gamma3} F^{1\gamma}{}_3  \nonumber\\
 & - g^{ab} \left(\partial_a \sigma_2 \right) \left(\partial_b\sigma_1 \right),
 \end{align}
 \begin{align}
 \hat R_{\bar 6 \bar 6}= & -\hat R_{0\bar 6 0\bar 6} +\hat R_{1\bar 6 1\bar 6}+\hat R_{2\bar 6 2\bar 6}+\hat R_{3\bar 6 3\bar 6}+\hat R_{\bar 5 \bar 6 \bar 5 \bar 6} + \hat R_{\bar 6 \bar 6 \bar 6 \bar 6} \nonumber\\
 =&~ \left[\partial_b \left(\partial_a \sigma_2 \right) \right] e_0{}^a e_0{}^b +  \left(\partial_a \sigma_2 \right)\left(\partial_b \sigma_2 \right) e_0{}^a e_0{}^b +  \left(\partial_a \sigma_2 \right)\left(\partial_b e_0{}^a \right) e_0{}^b +  \left(\partial_a \sigma_2 \right)\left(\partial_b e_0{}^a \right) e_0{}^b \nonumber\\
& -\frac{1}{4} f_2^2 e^{2 \sigma_2} F^2_{\gamma0} F^{2\gamma}{}_0 -  \left[\partial_b \left(\partial_a \sigma_2 \right) \right] e_1{}^a e_1{}^b -  \left(\partial_a \sigma_2 \right)\left(\partial_b \sigma_2 \right) e_1{}^a e_1{}^b -  \left(\partial_a \sigma_2 \right)\left(\partial_b e_1{}^a \right) e_1{}^b \nonumber\\
&- \left(\partial_a \sigma_2 \right)\left(\partial_b e_1{}^a \right) e_1{}^b 
 +\frac{1}{4} f_2^2 e^{2 \sigma_2} F^2_{\gamma1} F^{2\gamma}{}_1- \left[\partial_b \left(\partial_a \sigma_2 \right) \right] e_2{}^a e_2{}^b -  \left(\partial_a \sigma_2 \right)\left(\partial_b \sigma_2 \right) e_2{}^a e_2{}^b \nonumber\\
 &- \left(\partial_a \sigma_2 \right)\left(\partial_b e_2{}^a \right) e_2{}^b - \left(\partial_a \sigma_2 \right)\left(\partial_b e_2{}^a \right) e_2{}^b  +\frac{1}{4} f_2^2 e^{2 \sigma_2} F^2_{\gamma2} F^{2\gamma}{}_2 -  \left[\partial_b \left(\partial_a \sigma_2 \right) \right] e_3{}^a e_3{}^b \nonumber\\
 & -  \left(\partial_a \sigma_2 \right)\left(\partial_b \sigma_2 \right) e_3{}^a e_3{}^b - \left(\partial_a \sigma_2 \right)\left(\partial_b e_3{}^a \right) e_3{}^b -  \left(\partial_a \sigma_2 \right)\left(\partial_b e_3{}^a \right) e_3{}^b +\frac{1}{4} f_2^2 e^{2 \sigma_2} F^2_{\gamma3} F^{2\gamma}{}_3 \nonumber\\
 & - g^{ab} \left(\partial_a \sigma_2 \right) \left(\partial_b\sigma_1 \right).
\end{align}

\end{document}